\documentclass[aps,prd,twocolumn,showpacs,longbibliography]{revtex4-1}

\usepackage{amsmath,amssymb,amsfonts,amsthm}
\usepackage{dsfont}
\usepackage{CJK}
\usepackage{CJKpunct}

\usepackage{graphicx}
\usepackage{xcolor}

\usepackage[CJKbookmarks, pdftex, bookmarksnumbered, bookmarksopen, colorlinks, citecolor=blue, linkcolor=blue]{hyperref}

\newcommand{\Rl}[1]{\mathrm{Re}(#1)}                
\newcommand{\p}{\partial}                           
\newcommand{\dd}{\mathrm{d}}                        
\newcommand{\im}{\mathrm{i}}                        
\newcommand{\ex}{\mathrm{e}}                        
\newcommand{\tr}{\mathrm{tr}}                       
\newcommand{\ep}{\varepsilon}                       
\newcommand{\epn}{\epsilon}                         
\newcommand{\ket}[1]{|{#1}\rangle}                  
\newcommand{\bra}[1]{\langle{#1}|}                  
\newcommand{\braket}[2]{\langle{#1}|{#2}\rangle}    

\newcommand{\R}{\mathbb{R}}                         
\newcommand{\C}{\mathbb{C}}                         
\newcommand{\D}{\mathcal{D}}                        
\newcommand{\la}[1]{\mathfrak{{#1}}}                
\newcommand{\irp}[1]{\text{Irrep}^{(#1)}}           
\newcommand{\sgn}{\mathrm{sgn}}                     

\CJKtilde   

\newtheorem{Theorem}{Theorem}[section]
\newtheorem{Definition}{Definition}[section]
\newtheorem{Lemma}[Theorem]{Lemma}

\newtheorem{Proposition}[Theorem]{Proposition}
\newcommand{\startproof}{\textbf{Proof:\ \ }}

\begin{document}
  \title{Asymptotics of Spin Foam Amplitude on Simplicial Manifold: Euclidean Theory}
  \author{Muxin Han}
  \email{Muxin.Han@cpt.univ-mrs.fr}
  \affiliation{Centre de Physique Th\'eorique de Luminy, Case 907, F-13288 Marseille, France}
  \thanks{Unit\'e mixte de recherche du CNRS et des Universit\'es de Provence, de la M\'editerran\'ee et du Sud; affili\'e \`a la FRUMAN.}
  \author{Mingyi Zhang}
  \email{Mingyi.Zhang@cpt.univ-mrs.fr}
  \affiliation{Centre de Physique Th\'eorique de Luminy, Case 907, F-13288 Marseille, France}
  \thanks{Unit\'e mixte de recherche du CNRS et des Universit\'es de Provence, de la M\'editerran\'ee et du Sud; affili\'e \`a la FRUMAN.}
  \date{\today}

  \begin{abstract}
  We study the large-$j$ asymptotics of the Euclidean EPRL/FK spin foam amplitude on a 4d simplicial complex with arbitrary number of simplices. We show that for a critical configuration $\{j_f, g_{ve}, n_{ef}\}$ in general, there exists a partition of the simplicial complex into three regions: Non-degenerate region, Type-A degenerate region and Type-B degenerate region. On both the non-degenerate and Type-A degenerate regions, the critical configuration implies a non-degenerate Euclidean geometry, while on the Type-B degenerate region, the critical configuration implies a vector geometry. Furthermore we can split the Non-degenerate and Type-A regions into sub-complexes according to the sign of Euclidean oriented 4-simplex volume. On each sub-complex, the spin foam amplitude at critical configuration gives a Regge action that contains a sign factor $\mathrm{sgn}(V_4(v))$ of the oriented 4-simplices volume. Therefore the Regge action reproduced here can be viewed as a discretized Palatini action with on-shell connection. The asymptotic formula of the spin foam amplitude is given by a sum of the amplitudes evaluated at all possible critical configurations, which are the products of the amplitudes associated to different type of geometries.
   \end{abstract}
  \maketitle
  \begin{widetext}
  \tableofcontents
  \end{widetext}

  \newpage

  \section{Introduction}
  Loop Quantum Gravity (LQG) is an attempt to make a background independent, non-perturbative
quantization of 4-dimensional General Relativity (GR) -- for reviews, see \cite{Rovelli:2011eq,Rovelli:2010bf,Rovelli:QG,Perez:2003vx,Thiemann:QG,Han:2005km}. It is
inspired by the classical formulation of GR as a dynamical theory of connections. Starting
from this formulation, the kinematics of LQG is well-studied and results in a successful kinematical
framework (see the corresponding chapters in the books \cite{Rovelli:QG,Thiemann:QG}), which is also unique in a
certain sense. However, the framework of the dynamics in LQG is still largely open so
far. There are two main approaches to the dynamics of LQG, they are (1) the Operator formalism of
LQG, which follows the spirit of Dirac quantization or reduced phase space quantization of constrained dynamical system, and performs a canonical quantization of GR \cite{Thiemann96b,Han:2005zj,Giesel:2006uj,Giesel:2006uk,Giesel:2006um,Giesel:2007wn}; (2) the covariant formulation of LQG, which is currently understood in terms of the spin foam models \cite{Rovelli:2011eq,Barrett:1997gw,Barrett:1999qw,Engle:2007qf,Engle:2007uq,Engle:2007wy,Pereira:2007nh,Freidel:2007py,Livine:2007vk,Livine:2007ya,Han:2010rb}. The
relation between these two approaches is well-understood in the case of 3d quantum gravity
\cite{Noui:2004iy}, while for 4d quantum gravity, the situation is much more complicated and there are
some attempts \cite{Engle:2009ba,Han:2009aw,Han:2009ay,Han:2009az,Han:2009bb} for relating these two approaches.

The present article is concerning the framework of spin foam models. The current spin foam models for quantum gravity are mostly inspired by the 4-dimensional Plebanski formulation of GR \cite{Plebanski:1977zz,Reisenberger:1996pu,DePietri:1998mb} (or Plebanski-Holst formulation by including the Barbero-Immirzi parameter $\gamma$), which is a BF theory constrained by the condition that the $B$ field should be ``simple'' i.e. there is a tetrad field $e^I$ such that $B=\star(e\wedge e)$. Currently one of the successful spin foam models is the EPRL/FK model defined in \cite{Engle:2007qf,Engle:2007uq,Engle:2007wy,Pereira:2007nh,Freidel:2007py}, whose implementation of simplicity constraint is understood in the sense of \cite{Ding:2009jq,Ding:2010ye,Ding:2010fw}. The EPRL vertex amplitude is shown to reproduce the classical discrete GR in the large-$j$ asymptotics \cite{Barrett:2009gg,Barrett:2009mw}. Recently, The fermion coupling is included in the framework of EPRL spin foam model \cite{Bianchi:2010bn,Han:2011as}, and a q-deformed EPRL spin foam model is defined and gives discrete GR with cosmological constant in the large-$j$ asymptotics \cite{Han:2010pz,Fairbairn:2010cp,Ding:2011hp,Han:2011aa}.

The semiclassical behavior of the spin foam models is currently understood in terms of the \emph{large-$j$ asymptotics} of the spin foam amplitude, i.e. if we consider a spin foam model as a state-sum
\begin{equation}
  Z(\Delta)=\sum_{j_f}\mu(j_f)Z_{j_f}(\Delta)
\end{equation}
where $\mu(j_f)$ is a measure. We are investigating the asymptotic behavior of the (partial-)amplitude $Z_{j_f}$ as all the spins $j_f$ are taken to be large uniformly. The area spectrum in LQG is given approximately by $A_f=\gamma j_fl_p^2$, so the semiclassical limit of spin foam models is argued to be achieved by taking $l_p^2\to0$ while keeping the area $A_f$ comparable to the physical area, which leads to $j_f\to\infty$ uniformly as $\gamma$ is a fixed Barbero-Immirzi parameter. There is another argument relating the large-$j$ asymptotics of the spin foam amplitude to the semiclassical limit, by imposing the semiclassical boundary state to the vertex amplitude \cite{Bianchi:2010mw}. Mathematically the asymptotic problem is posed by making a uniform scaling for the spins $j_f\mapsto\lambda j_f$, and studying the asymptotic behavior of the amplitude $Z_{\lambda j_f}(\Delta)$ as $\lambda\to\infty$.

There was various investigations for the large-$j$ asymptotics of the spin foam models. The asymptotics of the Barrett-Crane vertex amplitude (10$j$-symbol) was studied in \cite{Barrett:2002ur}, which showed that the degenerate configurations in Barrett-Crane model were non-oscillatory, but dominant. The large-$j$ asymptotics of the FK model was studied in \cite{Conrady:SL2008}, concerning the non-degenerate Riemannian geometry, in the case of a simplicial manifold without boundary. The large-$j$ asymptotics of the EPRL model was initially studied in \cite{Barrett:2009gg,Barrett:2009mw} in both Euclidean and Lorentzian cases, where the analysis was confined into a single 4-simplex amplitude (EPRL vertex amplitude). It was shown that the asymptotics of the vertex amplitude is mainly a Cosine of the Regge action in a 4-simplex if the boundary data admits a non-degenerate 4-simplex geometry, and the asymptotics is non-oscillatory if the boundary data does not admit a non-degenerate 4-simplex geometry. There were also recent works to find the Regge gravity from the Euclidean/Lorentzian spinfoam amplitude on a simplicial complex via a certain ``double scaling limit'' \cite{Magliaro:Re2011,Magliaro:2011zz}.

The work presented here analyzes the large-$j$ asymptotic analysis of the Euclidean EPRL spin foam amplitude to the general situation with an 4d simplicial manifold with or without boundary, with an arbitrary number of simplices. The analysis for the Lorentzian EPRL model is presented in \cite{Han:2011AsLorentz}. The asymptotic behavior of the spin foam amplitude is determined by the stationary configurations of the ``spin foam action'', and is given by a sum of the amplitudes evaluated at the stationary configurations. Therefore the large-$j$ asymptotics is clarified as long as we find all the critical configurations and clarify their geometrical implications. Here for the Euclidean EPRL spin foam amplitude, a critical configuration in general is given by the data $\{j_f, g_{ve},n_{ef}\}$ that solves the equations of motion, where $j_f$ is an SU(2) spin assigned to each triangle, $g_{ve}$ is a SO(4) group variable, and $n_{ef}\in S^2$. In this work we show that given a general critical configuration, there exists a partition of the simplicial complex $\Delta$ which contains three types of regions: Non-degenerate region, Type-A (BF) degenerate region and Type-B (vector geometry) degenerate region. All of the three regions are simplicial sub-complexes with boundaries, and may be disconnected regions. The critical configuration implies different types of geometries in different types of regions:
\begin{itemize}
\item The critical configuration restricted in Non-degenerate region is non-degenerate in our definition of degeneracy. It implies a non-degenerate discrete Euclidean geometry on the simplicial sub-complex.

\item The critical configuration restricted in Type-A region is degenerate of Type-A in our definition of degeneracy. But it still implies a non-degenerate discrete Euclidean geometry on the simplicial sub-complex.

\item The critical configuration restricted in Type-B region is degenerate of Type-B in our definition of degeneracy. It implies a vector geometry on the simplicial sub-complex.
\end{itemize}

With the critical configuration, we further make a subdivision of the Non-degenerate and Type-A regions into sub-complexes (with boundary) according to their Euclidean oriented 4-volume $V_4(v)$ of the 4-simplices, such that $\mathrm{sgn}(V_4(v))$ is a constant sign on each sub-complex. Then in the each sub-complex, the spin foam amplitude at the critical configuration gives an exponential of Regge action in Euclidean signature. However we emphasize that the Regge action reproduced here contains a sign factor $\mathrm{sgn}(V_4(v))$ related to the oriented 4-volume of the 4-simplices, i.e.
\begin{equation}
S=\mathrm{sgn}(V_4)\sum_{\text{Internal}\ f} A_f\Theta_f+\mathrm{sgn}(V_4)\sum_{\text{Boundary}\ f} A_f\Theta_f^B
\end{equation}
where $A_f$ is the area of the triangle $f$ and $\Theta_f,\Theta_f^B$ are deficit angle and dihedral angle respectively. Recall that the Regge action without $\mathrm{sgn}(V_4)$ is a discretization of Einstein-Hilbert action of GR. Therefore the Regge action reproduced here is actually a discrete Palatini action with the on-shell connection (compatible with the tetrad).

The asymptotic formula of the spin foam amplitude is given by a sum of the amplitudes evaluated at all possible stationary configurations, which are the products of the amplitudes associated to different type of geometries.

Additionally, we also show in Section \ref{sec:parity} that given a spin foam amplitude $Z_{j_f}(\Delta)$ with the spin configuration $j_f$, any pair of the non-degenerate critical configurations associated with $j_f$ are related each other by a \emph{local} parity transformation. The parity transformation is the one studied in \cite{Barrett:2009gg,Barrett:2009mw} in the case of a single 4-simplex. A similar result holds for any pair of the degenerate configuration of Type-A associated with $j_f$, since it still relates to non-degenerate Euclidean geometry.

The article is organized as follow: In Section\ref{sec:Def}, we give a brief review of EPRL/FK spin foam amplitude and write the transition amplitude in a path integral form. In Section\ref{sec:SemiCd}, we discuss the semiclassical limit we are considering. A detail discussion of classical discrete geometry on a simplicial complex is in Section\ref{sec:DcG}. The non-degenerate critical configuration is discussed in detail in Sections \ref{sec:EoM},\ref{sec:SemiGV},\ref{sec:AsympND}, and \ref{sec:parity}. The degenerate Type-A, Type-B configurations are discussed in Section\ref{sec:AsympD}. In Section\ref{sec:AsymptoRegge} we give the asymptotics of the spin foam amplitude as a sum over all possible critical configurations.

  \section{Spin Foam Amplitude}\label{sec:Def}
  In this section we briefly review the definition of the Euclidean EPRL spin foam amplitude. We denote $\Delta$ as a simplicial complex and $\Delta^*$ as its dual. The building blocks in $\Delta$ are 4-simplices $\sigma_v$, tetrahedrons $t_e$ and triangles $f$. The corresponding dual building blocks in $\Delta^*$ are vertices $v$, edges $e$ and faces $f$, respectively. We identify the notations of triangle and face, because there is a 1-to-1 correspondence between the triangles in $\Delta$ and a dual face in $\Delta^*$. The orientation of $\Delta^*$ is determined by the orientation of $e$ and $f$. We call $\Delta^*$ is \emph{oriented} as long as the orientations of $e$ and $f$ are chosen.

  For defining spin foam model, we introduce more structures to $\Delta^*$. For each internal edge $e$ with $\p e=(vv')$, we cut it into two half-edges $(ve)$ and $(ev')$ at the middle point of $e$ (we denote the middle point of $e$ also by $e$). The orientation of the half-edge are always from $e$ to $v$. We associate a group element $g_{ve}\in$SO(4) to each half-edge $(ve)$, and associate a irreducible representation $\irp{j^+,j^-}$[SO(4)] to each face. At each edge $e$ we associate an SU(2) coherent intertwiner with the resolution of identity \cite{Livine:2007vk}
  \begin{equation}
    \mathds{1}_{H_i}=\int\prod_{f\in e} \dd^2 n_{ef}|\ket{j_f,n_{ef}}\bra{j_f,n_{ef}}|
  \end{equation}
  As proved in \cite{Conrady:2009px}, the above integration is essentially over the constraint surface of closure constraint $\sum_{f\subset t_e}j_f n_{ef}=0$. It means the labels of coherent intertwiner $j_f$ and $n_{ef}$ have geometrical interpretation in the quantum level as a tetrahedron. With the coherent intertwiner, we impose closure constraint to the spin foam amplitude and associate $e$ with a geometrical tetrahedron $t_e$. In our following discussion we assume all the tetrahedrons $t_e$ are non-degenerate. $\gamma j_f$ is the area of the triangle $f$. In the definition of the spin foam amplitude as a state sum, we only sum over the spins with $\sum_{f\subset t_e}\epsilon_f j_f\neq0$, for all $\epsilon_f=\pm1$ and for all tetrahedrons $t_e$, so that all the geometrical tetrahedrons are non-degenerate. $n_{ef}$ stands for the unit 3-vector normal to the triangle $f$ of the tetrahedron $t_e$. There is a unit 4-vector $u_e=(1,0,0,0)$ orthogonal to all $n_{ef}$. For each edge $e$ connecting to the boundary and connecting to an internal vertex $v$, we regard it as a half edge and associate it with $g_{ve}\in \text{SO(4)}$. We associate the edge $e$ with boundary intertwiners $|\ket{j_f,n_{ef}}$ (or $\bra{j_f,n_{ef}}|$) with boundary data $j_f,n_{ef}$.

  Based on the definitions and notations above we can write down the spin foam model. The definition of EPRL spin foam model we can find in many articles e.g. \cite{Engle:2007wy}\cite{Bianchi:2010mw}\cite{Rovelli:2010vv}. Usually the spin foam amplitude is written in terms of a product of vertex amplitudes $A_v$ and face amplitudes $A_f$,, followed by the sums/integrations over the variables $(j_f,g_{ve},n_{ef})$
  \begin{equation}
    Z =\sum_{j_{f}}\int \prod_{\left( ve\right) }\dd g_{ve}\int \prod_{\left(ef\right) }\dd n_{ef}\prod_{f}A_{f}\prod_{v}A_{v}\left(g_{ve},j_{f},n_{ef}\right).
  \end{equation}

  In the following we are going to write the spin foam amplitude into a ``path integration'' form as $\int D\mu\ e^S$, i.e. we can express the spin foam amplitude as the follows
  \begin{equation}
    Z(j_{f_e})=\sum_{j_{f_i}}\prod_{f}\mu \left( j_{f}\right) \int \prod_{\left( ve\right)}\dd g_{ve}\int \prod_{\left( ef_i\right)}\dd n_{ef_i}\ex^{\sum_{f}S_{f}}
  \end{equation}
  where $f_e$ and $f_i$ mean boundary and internal faces respectively, and
  \begin{equation}\label{eq:actionf}
    S_{f}=\sum_{v\in f}\ln \left\langle j_{f},n_{ef}\left\vert Y^{\dag}g_{ev}g_{ve^{\prime }}Y\right\vert j_{f},n_{e^{\prime }f}\right\rangle
  \end{equation}
$S=\sum_fS_f$ is an ``spin foam action'' for the path integral. It turns out that the critical point of the spin foam action determines the asymptotic behavior of the spin foam amplitude as $j\to\infty$. In the above result we have already absorbed the SU(2) integration in coherent intertwiner into the integration of $g_{ve}$. The similar formulas can be found in \cite{Conrady:2008ea,Conrady:SL2008,Magliaro:2011qm,Magliaro:Re2011}. Here we use notation $g_{ev}\equiv g_{ve}^{-1}$. $Y$ is a projector $Y:\irp{j}$[SU(2)]$\rightarrow\irp{j^+,j^-}$[SO(4)]. Using this projector we can totally decompose SO(4) group into its self-dual $g^+$ and anti-self-dual $g^-$ parts where $g^+, g^-\in \mathrm{SU(2)}, \forall g\in\mathrm{SO(4)}, g=g^-(g^+)^{-1}$ and insert the simplicity condition $j^{\pm}=(1\pm\gamma)/2$. The above result works for the case with the Barbero-Immirzi parameter $\gamma<1$. The case with $\gamma>1$ will included in the discussion starting from Section\ref{sec:EoM}.

Moreover, the spin foam action $S$ can be written in the following form
  \begin{equation}\label{eq:actionE}
    \begin{split}
      S &=\sum_f\sum_{v\in f}\ln  \left\langle j_{f},n_{ef}\left\vert Y^{\dag }g_{ev}g_{ve^{\prime }}Y\right\vert j_{f},n_{e^{\prime}f}\right\rangle\\
      &=\sum_f\sum_{v\in f}\sum_{\pm}2j_{f}^{\pm }\ln \left\langle n_{ef}\left\vert g_{ev}^{\pm }g_{ve^{\prime }}^{\pm }\right\vert n_{e^{\prime }f}\right\rangle
    \end{split}
  \end{equation}
  where $|n\rangle$ is a coherent state in the fundamental representation. It is normalized $\braket{n}{n}=1$ and can be represented by a spinor $|n\rangle=\xi_{\alpha}=(z_0,z_1)$, where $z_0, z_1\in \C$. We can identify the spinor with a unit 3-vector $\mathbf{n}$, where the component of $\mathbf{n}$ is defined as the follows
  \begin{equation}\label{eq:defCS}
    |n\rangle \langle n|=\xi_{\alpha}\bar{\xi}_{\dot{\alpha}}=\frac{1}{2}(\delta_{\alpha\dot{\alpha}}+n_i\sigma^i_{\alpha\dot{\alpha}}).
  \end{equation}

  The spin foam action $S$ is written as a sum of the ``face action'' $S_f$ over all the faces. Here in this paper, we are going to compare the spin foam action at the critical point with the Regge action
  \begin{equation}
    S_{R}=\sum_{f}A_{f}\Theta _{f}
  \end{equation}
  where $A_{f}$ is the area of the triangle $t$ dual to the face $f$, $\Theta _{f}$ is the deficit angle in $f$.

  \section{Semiclassical considerations}\label{sec:SemiCd}
  In this section we pose the asymptotic problem towards clarifying the semiclassical limit of the EPRL spin foam amplitude.

It is argued that the semiclassical limit in spin foam formulation is achieved by taking $l_p^2\to0$ while keeping the physical area $A_f=\gamma j_fl_p^2$ fixed, which implies that $j\rightarrow \infty $ as the limit to obtain the semiclassical approximation. Mathematically we rescale all the internal and boundary $j$s with a uniform scaling parameter $\lambda$. Then the large-$j$ limit is taken by sending $\lambda\rightarrow\infty$. Here we emphasis that the semiclassical limit is different from the continuum limit, as discussed by Rovelli in \cite{Rovelli:2011mf}. The continuum limit of the theory (even within its semiclassical regime) is out of the scope of the present paper. With the large-$j$ limit taken here, we will obtain in some sense the classical GR truncated on the simplicial complex. But to achieve a continuum formulation is out of the scope of this paper.

  As discussed in \cite{Conrady:SL2008,Barrett:2009gg,Barrett:2009mw}, the asymptotic behavior of spin foam amplitude is determined by the critical points of the spin foam action $S$, i.e. the stationary phase points of $S$ satisfying $\Rl{S}=0$. The amplitude at the configurations which do not satisfy these two conditions are all exponentially suppressed in the large-$j$ limit.

  Here we write a spin foam amplitude as
 \begin{equation}
    Z(\Delta)=\sum_{j_f}\mu(j_f)Z_{j_f}(\Delta)
  \end{equation}
we are studying the asymptotic behavior of the (partial-)amplitude $Z_{\lambda j_f}(\Delta)$ as $\lambda\to\infty$. We do not study the stationary phase with respect to spin $j$s, and expect the sum over spin $j$s should become the sum over all the classical areas once the large-$j$ limit is taken. We will clarify the geometric meaning of the face spins in the large-$j$ regime, i.e. $\gamma j_{f}$ is interpreted as the area $A_f$ of the triangle $f$. Thus in our calculation, the equation of motion we are considering is given by
  \begin{eqnarray}
    \Rl{S}&=&0 \label{eq:aR0}\\
    \delta_{g_{ve}}S&=&0, \label{eq:ag0}\\
    \delta_{n_{ef}}S&=&0. \label{eq:an0}
      \end{eqnarray}
Under the large-$j$ limit we would like to compare the large-$j$ regime of the spin foam amplitude with path integral formulation of area Regge calculus
  \begin{equation}
    Z(j_{f_e})\sim\sum_{j_{f_i}\ \text{large}}\mu(j_f)\ex^{\im S_{\text{critical}}}\sim\int_{j_{f_i}}\D j_{f_i} \ex^{\im S_{\text{Regge}}}.
  \end{equation}
Note that there is the gluing of between 4-simplices imposed in the spin foam amplitude since there is only a single set of variables $(j_f,n_{ef})$ for each tetrahedron $t_e$. We will come back to this point later.

  \section{Discrete Geometry on Simplicial Complex}\label{sec:DcG}

  In this section we discuss the discrete geometry over a non-degenerate simplicial complex $\Delta$. The aim of this section is to give a collection of definitions and variables to describe the discrete Euclidean geometry on $\Delta$. These geometrical variables will be reconstructed from the critical configurations of spin foam amplitude in next sections. In the following we denote the 4d Euclidean vector space by $\mathbb{E}$.

  \subsection{The Orientation Structure of Simplicial Complex $\Delta$}
  In this subsection, we discuss the orientation structure of simplicial complex $\Delta$. The simplicial complex $\Delta$ is a triangulation of the space-time manifold $M$. Here we would like to review the definition of the orientation of $\Delta$, which is necessary to define e.g. oriented 4-volume for each 4-simplex. For convenience, we call the edge of triangle in $\Delta$ ``Segment'', denoted $l$, and the vertex of triangle called ``Point'', denoted $p$.

  \subsubsection{The orientation of $\Delta$}
\begin{figure}[htbp!]
    \centering
    \includegraphics[width=0.3\textwidth]{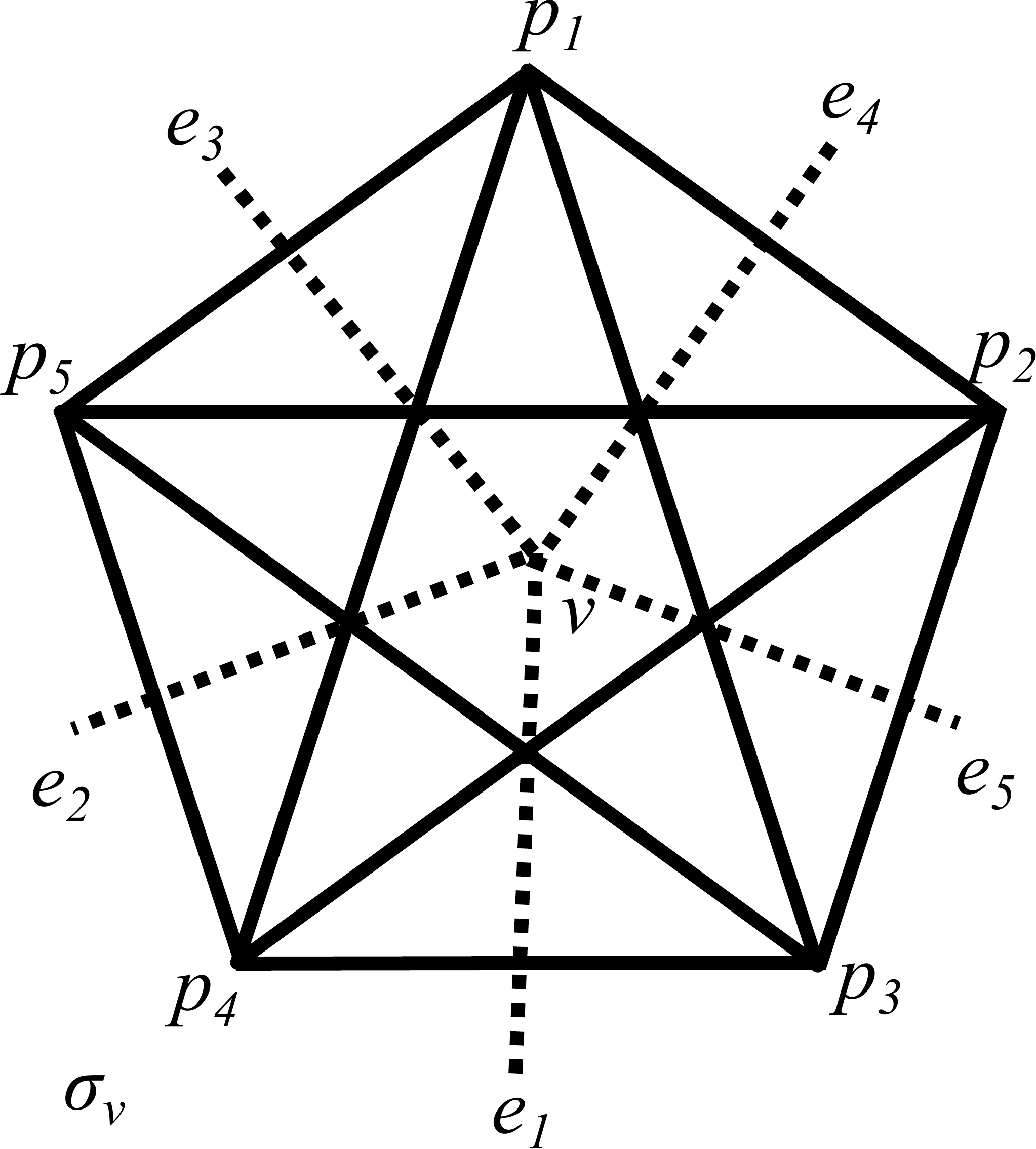}
    \caption{The duality between $p_i$ and $e_i$ in $\sigma_v$}\label{fig:4simplex}
  \end{figure}

  Before defining the orientation of the Simplicial Complex $\Delta$, we have to define the orientation of each 4-simplex $\sigma_v$ dual to $v$. The orientation of a 4-simplex $\sigma_v$ can be represented by its ordered 5 points, i.e. a tuple $[p_1,\cdots, p_5]$. Two orientations are opposite if the two tuples can be related by odd permutation, e.g. $[p_1,p_2,\cdots, p_5]=-[p_2,p_1,\cdots, p_5]$. Because in a 4-simplex $\sigma_v$, there are five points $p$ and five tetrahedrons $t_e$. We can make a duality between $p$ and $e$ if $p\cap t_e=\emptyset$, as shown in Fig.\ref{fig:4simplex}.
  Thus the orientation of $\sigma_v$ can also be denoted as $[e_1,\cdots, e_5]$. Since vertex $v\in\Delta^*$ is dual to $\sigma_v$, we can also say $[e_1,\cdots, e_5]$ is the orientation of vertex $v$.

  From the orientation of 4-simplex $\sigma_v$ we can induce the orientations of  tetrahedrons $t_e$, triangles $f$ and segments $l$ in $\sigma_v$. For example, we define the orientation of $t_{e_1}$ as
  \begin{equation}\label{eq:OrientT}
    [e_2, e_3, e_4, e_5]\leftarrow [e_2, e_1, e_3, e_4, e_5]
  \end{equation}
where $\leftarrow$ denotes the induction from the 4-simplex orientation to the orientation of $t_{e_1}$, by deleting the second entry of $[e_2, e_1, e_3, e_4, e_5]$. Similarly, the orientation of $f=t_{e_1}\cap t_{e_2}$ respect to $t_{e_1}$ can be defined as
  \begin{equation}\label{eq:OrientA}
    [e_3,e_4,e_5]\leftarrow [e_3, e_1, e_2, e_4, e_5]
  \end{equation}
 where the induction is given by deleting the second and third entries of $[e_3, e_1, e_2, e_4, e_5]$. The orientation of $l$ from point $p_4$ to point $p_5$ respect to $t_{e_1}$ can be defined as
  \begin{equation}
    [e_4,e_5]\leftarrow[e_3, e_1, e_2, e_4, e_5]
  \end{equation}
  For convenience, we will use a Levi-Civita symbol $\epn_{e_ie_je_ke_le_m}(v)$ to denote $[e_i, e_j,e_k, e_l, e_m]$ of $\sigma_v$ in the the following discussion.

  From Eq.(\ref{eq:OrientA}), we can find in $\sigma_v$, $\forall f=t_{e_i}\cap t_{e_j}$, the orientations of $f$ respect to $t_{e_i}$ and $t_{e_j}$ are opposite.

\begin{figure}[htbp!]
    \centering
    \includegraphics[width=0.4\textwidth]{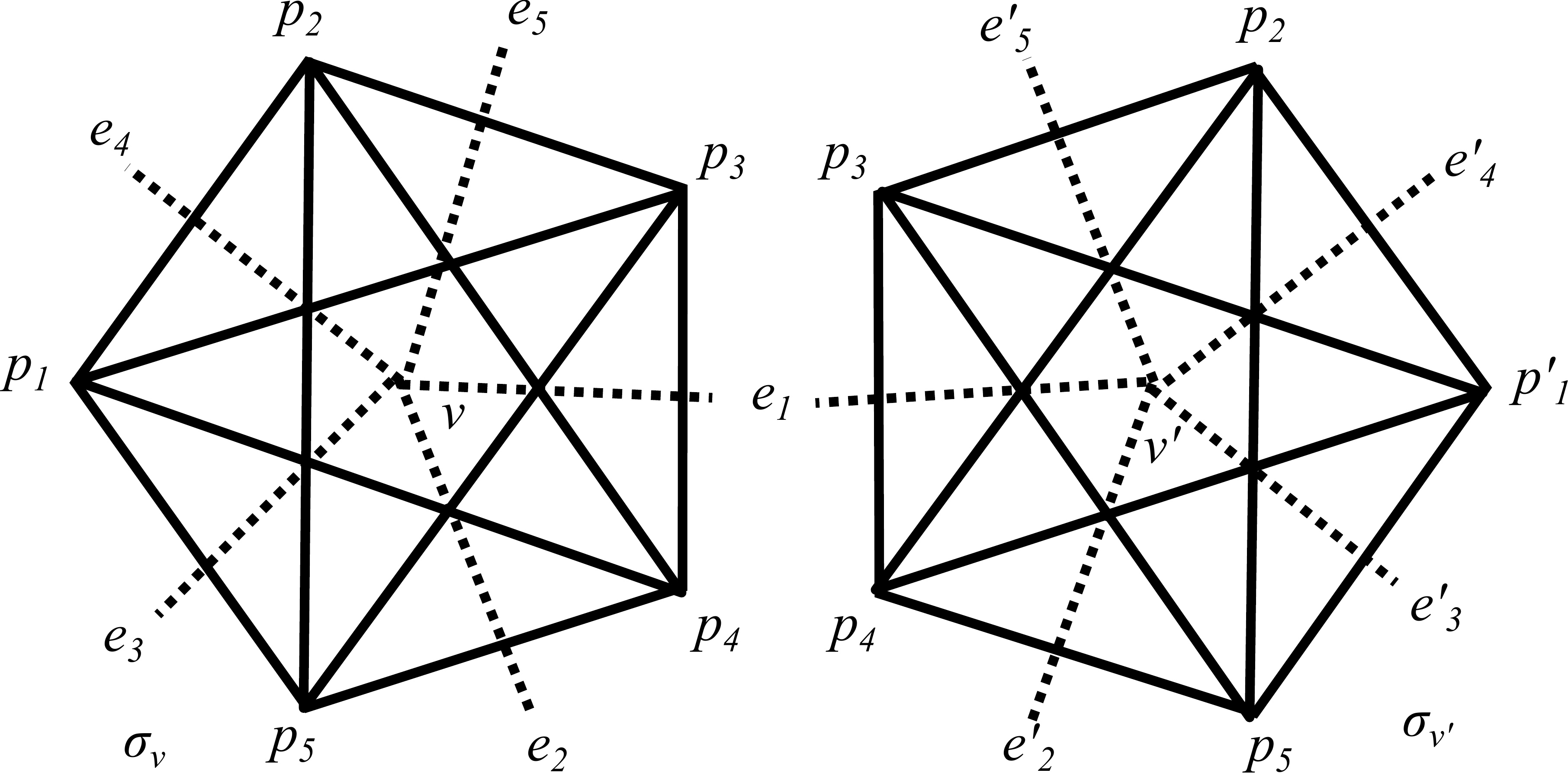}
    \caption{Two 4-simplexes $\sigma_v, \sigma_{v'}$ share tetrahedron $t_{e_1}$}\label{fig:Glue4simplex}
  \end{figure}

  We say that two neighboring 4-simplexes $\sigma_v$,$\sigma_{v'}$ in Fig.\ref{fig:Glue4simplex} are \emph{orientation consistent} if the orientations of the tetrahedron $t_{e_1}$ shared by them are opposite respecting to $\sigma_v$ and $\sigma_{v'}$, i.e. $[p_1,p_2,\cdots,p_5]=-[p'_1,p_2,\cdots,p_5]$ or $[e_1,e_2,\cdots,e_5]=-[e_1,e'_2,\cdots,e'_5]$. It means
  \begin{equation}\label{eq:OrientConsist}
    \epn_{e_1e_2e_3e_4e_5}(v)=-\epn_{e_1e'_2e'_3e'_4e'_5}(v')
  \end{equation}

  From this we can see that the orientations of triangles $f\in t_{e_1}$ and segments $l\in f$ are opposite respecting to $\sigma_v$ and $\sigma_{v'}$ as
  \begin{eqnarray}
    \epn_{e_ie_je_k}(v)&=&-\epn_{e'_ie'_je'_k}(v'), \quad \forall i,j,k\neq 1\\
    \epn_{e_ie_j}(v)&=&-\epn_{e'_ie'_j}(v'), \quad \forall i,j\neq 1
  \end{eqnarray}

  We call a given simplicial complex $\Delta$ (or $\Delta^*$) is \emph{global oriented} if any two neighboring 4-simplexes (or vertices)in $\Delta$ are orientation consistent. In the following discussion in this section, we assume the simplicial complex $\Delta$ is global oriented.

  \subsubsection{Space-time orientation}

We assume the simplicial complex $\Delta$ is a discretization of a manifold $M$ with a global orientation. Therefore we can define an \emph{oriented} orthonormal frame bundle $e_\mu^I$, where all the orthonormal frames are right-handed with respect to the global orientation, or $\sgn\det(e)$ is a constant sign on the manifold $M$. The oriented orthonormal frame bundle has the structure of a principle fiber bundle with the structure group SO(4) in Euclidean signature (or SO(1,3) in Lorentzian signature).

Now we give a discrete analogue of a global space-time orientation on a simplicial manifold. Given a simplicial complex $\Delta$, we assign a reference frame $\{e(v)\}$ in each 4-simplex $\sigma_v$. We assume for any two frames $\{e(v)\}$ and $\{e(v')\}$ at two different 4-simplices,
  \begin{equation}\label{eq:eee}
    \mathrm{sgn}\det e(v)=\sgn\det e(v')
  \end{equation}
Then the two reference frame between two neighboring simplexes are related by an SO(4) transformation. Then the frames located in different 4-simplices constitute a discrete analogue of {oriented} orthonormal frame bundle on the simplicial manifold. The SO(4) transformation relating the two frames in different 4-simplices are the discrete spin connection.

Moreover, in next subsections, we will show that, if there exists a discrete analogue of the {oriented} orthonormal frame bundle on $\Delta$, i.e. there are frames assigned in the 4-simplices satisfying $\mathrm{sgn}\det e(v)=\sgn\det e(v')$ and being related with each other by SO(4) transformations, then for the oriented volume of 4-simplex, the sign $\sgn(V_4(v))$ is a constant, with a consistent orientation on the simplicial complex $\Delta$.i.e.
  \begin{equation}\label{eq:VeV}
     \mathrm{sgn}V_4(v)=\mathrm{sgn}V_4(v'), \ \ \ \forall\ v,v'.
  \end{equation}
See the next subsection for the definition of $V_4(v)$.

  \subsection{Discrete Geometry in a 4-simplex}
Given a simplicial complex $\Delta$, we can define a collection of geometric variables to describe the discrete geometry on a simplicial manifold.

  \begin{Definition}[Segment Vector $E_l(v)$]
    A segment vector $E^I_l(v)$ is a 4-vector in the tangent space of $v\in\Delta^*$, associated with the oriented segment $l\in\Delta$. The modulus of $E^I_l(v)$ is the length of $l$. $E_l(v)$ should satisfy the following properties:
  \begin{itemize}
    \item[Inverse:] When the orientation of $l$ is inversed,
    \begin{equation}\label{eq:inverse}
      E^I_{-l}(v)=-E^I_l(v)
    \end{equation}
    \item[Close:] $\forall f\in\Delta$, if its boundary $l_1,~l_2,~l_3$ orientations are consistent, then
    \begin{equation}\label{eq:Close}
      E^I_{l_1}(v)+E^I_{l_2}(v)+E^I_{l_3}(v)=0
    \end{equation}
    \item[Gluing:] If $\sigma_v\cap\sigma_{v'}=t_e$, $\forall f\in\p t_e,~ l,l' \in\p f$
    \begin{equation}\label{eq:gluing}
      \delta_{IJ}E_{l}^I(v)E_{l'}^J(v)=\delta_{IJ}E_{l}^I(v')E_{l'}^J(v').
    \end{equation}
  \end{itemize}
  \end{Definition}

  As in \cite{Conrady:SL2008}\cite{Hamber:QG-Path}, the segment vectors $E_l^I(v)$ are natural (co-)frames for discrete geometry. The discrete version of the metric is defined by
  \begin{equation}
    g_{l_1l_2}(v)= \delta_{IJ}E_{l_1}^I(v)E_{l_2}^J(v)
  \end{equation}
  where $l_1$ and $l_2$ are in the same triangle. It is independent of the choice of $v$ because of the gluing property Eq.(\ref{eq:gluing}).

  For the case we are considering, we also need to define the non-degeneracy of $E_l^I(v)$.
  \begin{Definition}[Non-degeneracy]
    We call $E_l^I(v)$ non-degenerate if $\forall p,l\in\sigma,~ l\cap p\neq\emptyset,~s.t.~E_l^I(v)$ spans a 4-dimensional vector space.
  \end{Definition}

  Because a segment $l$ can be denoted by its end-points $l=[p_lp'_l]$, it can be also denoted by the edges dual to its end-points, i.e. $l=[p_lp'_l]=[ee']$, where $[ee']=-[e'e]$. Thus we can also write $E_l^I(v)$ as $E_{ee'}^I(v)$. The direction of $E_{ee'}(v)$ is from $e$ to $e'$. Then the Inverse and Close properties turn to
  \begin{eqnarray}
    &&\ \ \ \ \ \ \ \ \ E_{ee'}^I(v)=-E_{e'e}^I(v),\nonumber\\
    &&\quad E_{e_1e_2}^I(v)+E_{e_2e_3}^I(v)+E_{e_3e_1}^I(v)=0
  \end{eqnarray}
  In the following we use both conventions, according to the convenience of the context.

We choose a consistent orientation of all the 4-simplices of $\Delta$. Then for each 4-simplex $\sigma_v\in\Delta$, we can define the oriented 4-volume $V_4(v)$ of $\sigma_v$
  \begin{equation}\label{eq:V4}
    \begin{split}
      V_4(v)&\equiv\det\left(E_{e_2e_1}(v),E_{e_3e_1}(v),E_{e_4e_1}(v),E_{e_5e_1}(v)\right)\\
    &=\frac{1}{4!}\sum_{j,k,l,m}(\epn^{ijklm}\epn_{IJKL}E_{e_je_i}^IE_{e_ke_i}^JE_{e_le_i}^KE_{e_me_i}^L)(v)
    \end{split}
  \end{equation}
  which is independent of the index $i$ by Eq.(\ref{eq:Close}). Here $\epn^{ijklm}$ and $\epn_{IJKL}$ are Levi-Civita symbol, with $\epn^{ijklm}=\epn_{ijklm}$ and $\epn_{IJKL}=\epn^{IJKL}$. We define five 4-vectors $U^e(v)$ orthogonal to $t_e$ by
  \begin{equation}\label{eq:U}
    U_I^{e_k}(v)\equiv\frac{1}{3!V_{4}\left( v\right) }\sum_{l,m,n}(\epsilon ^{jklmn}\epsilon_{IJKL}E_{e_{l}e_{j}}^{J} E_{e_{m}e_{j}}^{K}E_{e_{n}e_{j}}^{L})\left( v\right)
  \end{equation}
  We call them \emph{frame vectors}. Using the above definition and Eq.(\ref{eq:inverse}), Eq.(\ref{eq:Close}), we get
  \begin{equation}\label{eq:orthogonal}
    U_I^{e_i}(v)E_{e_je_k}^I(v)=\delta^i_j-\delta^i_k
  \end{equation}
  When we sum over the five $U^e(v)$ in Eq.(\ref{eq:orthogonal}), we obtain
  \begin{equation}
    \sum_i^5U_I^{e_i}(v)E_{e_je_k}^I(v)=\sum_i^5\delta^i_j-\sum_i^5\delta^i_k=0, \quad \forall e_j, e_k
  \end{equation}
  which implies the closure of $U^e(v)$ for each 4-simplex $\sigma_v$
  \begin{equation}\label{eq:closeU}
    \sum_i^5U_I^{e_i}(v)=0
  \end{equation}
  Eq.(\ref{eq:orthogonal}) and Eq.(\ref{eq:closeU}) show that in $v$, the 5 vectors are all out-pointing to the tetrahedrons from $\sigma_v$ up to a total reflection $U^e_I\to-U^e_I$. Also from Eq.(\ref{eq:orthogonal}) we obtain the following identities
  \begin{equation}
    V_4^{-1}(v)=\det \left(U^{e_2}(v),U^{e_3}(v),U^{e_4}(v),U^{e_5}(v)\right)
  \end{equation}
  \begin{equation}\label{eq:UtoE}
    E_{e_ke_j}^I(v)=\frac{V_4(v)}{3!}\sum_{l,m,n}\epn_{jklmn}\epn^{IJKL}U_J^{e_l}(v)U_K^{e_m}(v)U_L^{e_n}(v)
  \end{equation}
  \begin{equation}\label{eq:XUtoXE}
    \begin{split}
      &\quad V_4(v)(U^{e_i}(v)\wedge U^{e_j}(v))_{IJ}\\
      &=\frac{1}{2}\sum_{m,n}\epn^{kijmn}\epn_{IJKL}E_{e_me_k}^K(v)E_{e_ne_k}^L(v)
    \end{split}
  \end{equation}
  Here the last equation give us a way to construct area bivectors explicitly. In a 4-simplex $\sigma_v$, a triangle can be identified by the two tetrahedrons that share it, or by the three points of the triangle. We define two area bivectors of the triangle shared by $t_{e_1}$ and $t_{e_2}$, which are denoted by $A_{e_1e_2}$ and $A_{e_3e_4e_5}$. We define them in the following way
  \begin{eqnarray}\label{eq:DefA}
     A_{e_1e_2}(v)&=&\frac{1}{4}\sum_{m,n}\epn^{k12mn}(E_{e_me_k}(v)\wedge E_{e_ne_k}(v))\\
     A_{e_3e_4e_5}(v)&=&\frac{1}{2}(E_{e_3e_5}(v)\wedge E_{e_4e_5}(v))
  \end{eqnarray}
  The bivector $A_{e_1e_2}(v)$ depends on the orientation of the 4-simplex, while the bivector $A_{e_3e_4e_5}(v)$ is defined with an orientation of the triangle $f=[p_3,p_4,p_5]\equiv[e_3,e_4,e_5]$, which may or may not be the orientation induced from $\sigma_v$. We call $A_{e_ie_j}(v)$ \emph{oriented bivectors} and $A_{e_ie_je_k}(v)$ \emph{non-oriented bivectors}. Their relation is $A_{e_ie_je_k}(v)=\epn_{e_ie_je_ke_le_m}(v)A_{e_le_m}(v)$ (no sum in $e_l,e_m$).

  \subsection{Gluing Condition of Many 4-simplexes}
  Given a tetrahedron $t_{e_1}\in\Delta$ which is shared by $\sigma_v, \sigma_{v'}$, as in Fig.\ref{fig:Glue4simplex}, we consider the relation between $E_l^I(v)$ and $E_l^I(v')$ for $l\in t_{e_1}$. We define two unit normal vectors $\hat{U}_{e_1}^I(v)\equiv U_{e_1}^I(v)/|U_{e_1}(v)|$ and $\hat{U}_{e_1}^I(v')\equiv U_{e_1}^I(v')/|U_{e_1}(v')|$, where $U_{e_1}^I=\delta^{IJ}U^{e_1}_J$. From Eq.(\ref{eq:U}), we can find $\forall l\in t_{e_1}$
  \begin{equation}\label{eq:orth}
    \hat{U}_{e_1}^I(v)E_{lI}(v)=\hat{U}_{e_1}^I(v')E_{lI}(v')=0
  \end{equation}
  Thus for $l_1,l_2,l_3\in t_{e_1}$ but not in the same face, the vectors $E_{l_1},E_{l_2},E_{l_3},\hat{U}_{e_1}$ define two reference frames in both $\sigma_v$ and $\sigma_{v'}$. To satisfy Eq.(\ref{eq:eee}) we should have
  \begin{equation}\label{eq:sgn}
    \begin{split}
      &\text{sgn}~\det\left(E_{l_1}(v),E_{l_2}(v),E_{l_3}(v),\hat{U}_{e_1}(v)\right)\\     =&\text{sgn}~\det\left(E_{l_1}(v'),E_{l_2}(v'),E_{l_3}(v'),\hat{U}_{e_1}(v')\right)
    \end{split}
  \end{equation}
  where $E_{l_1}(v),E_{l_2}(v),E_{l_3}(v)$ and $E_{l_1}(v'),E_{l_2}(v'),E_{l_3}(v')$ span a three-dimensional subspace at $v$ and $v'$, respectively. Because of Eq.(\ref{eq:gluing}), Eq.(\ref{eq:orth}) and Eq.(\ref{eq:sgn}), there exists a unique SO(4) \cite{Conrady:SL2008,Han:2011AsLorentz} matrix $\Omega_{v'v}$ such that
  \begin{equation}\label{eq:EOEUOU}
    (\Omega_{v'v})^I_{~J}E^J_l(v)=-E^I_l(v'),\quad (\Omega_{v'v})^I_{~J}\hat{U}_{e_1}^J(v)=-\hat{U}^I_{e_1}(v')
  \end{equation}
  The minus in the first equation is because of the orientations of any segments $l\in t_{e_1}$ are opposite respecting to two neighboring 4-simplexes $\sigma_v$ and $\sigma_{v'}$. If $l=p_ip_j$, $E_l(v)=E_{p_ip_j}(v)$, $E_l(v')=E_{p_jp_i}(v')$, we can also rewrite it as
  \begin{equation}
    (\Omega_{v'v})^I_{~J}E^J_{p_ip_j}(v)=E^I_{p_ip_j}(v')
  \end{equation}
  The second equation is because
  \begin{equation}
    \begin{split}
      &\det\left(E_{l_1}(v),E_{l_2}(v),E_{l_3}(v),\hat{U}_{e_1}(v)\right)\\
      =&-\det\left(E_{l_1}(v'),E_{l_2}(v'),E_{l_3}(v'),\Omega_{v'v}\hat{U}_{e_1}(v)\right)\\   =&\det\left(E_{l_1}(v'),E_{l_2}(v'),E_{l_3}(v'),\hat{U}_{e_1}(v')\right)
    \end{split}
  \end{equation}
  The first equality is because $\det\Omega_{v'v}=1$. The second equality implies $\hat{U}_{e_1}(v)=-\Omega_{vv'}\hat{U}_{e_1}(v')$. We call $\Omega_{vv'}$ the \emph{spin connection} if it satisfies Eqs.(\ref{eq:sgn}), (\ref{eq:EOEUOU}).

\begin{figure}[htbp!]
    \centering
    \includegraphics[width=0.4\textwidth]{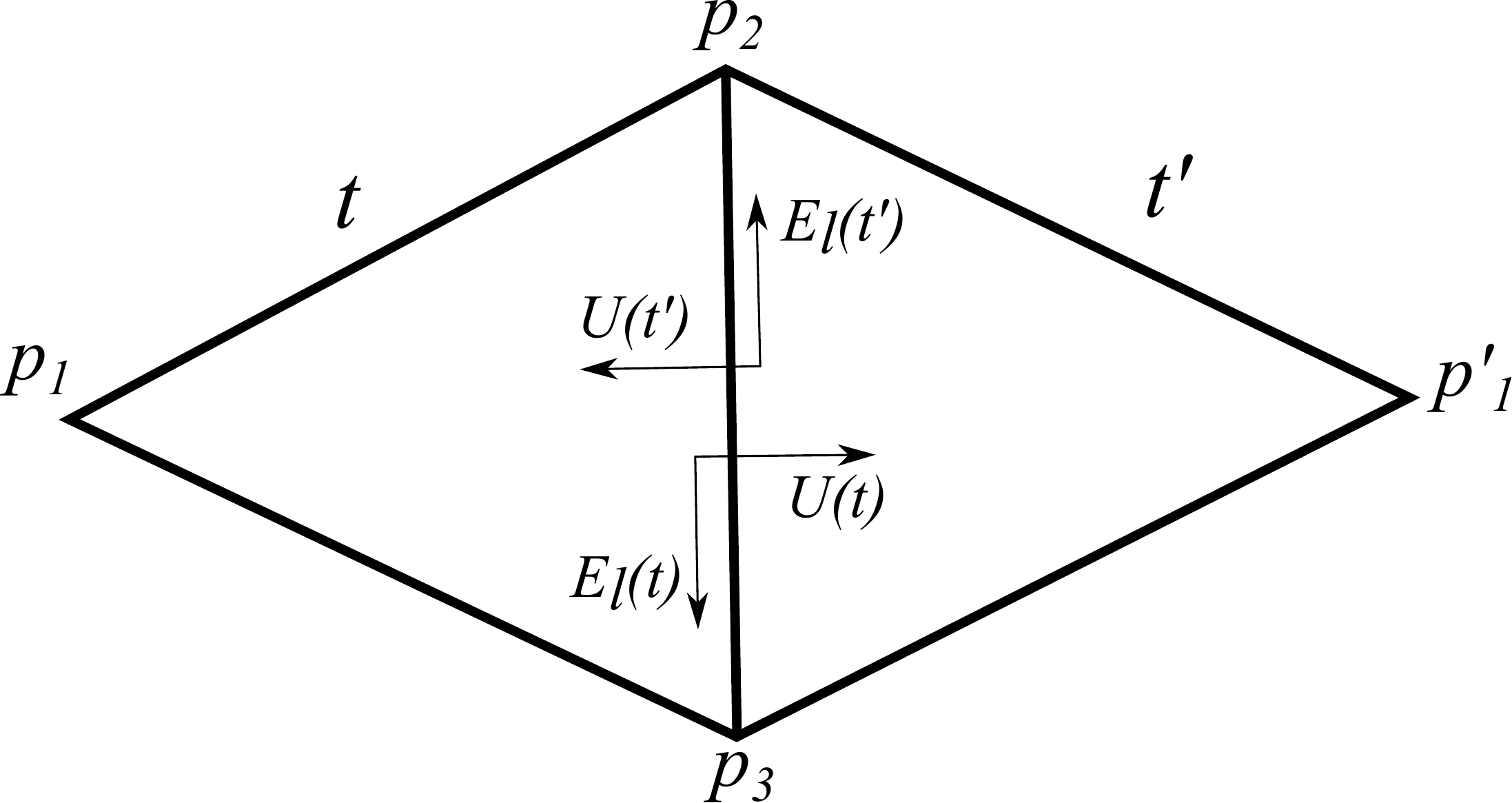}
    \caption{Two triangles $t, t'$ share segment $p_2p_3$}\label{fig:2d_eg}
  \end{figure}

  For an explanation of Eq.(\ref{eq:EOEUOU}), we see a 2-D example showing in Fig.\ref{fig:2d_eg}
  where two triangles share a segment $l=p_2p_3$. Because of the orientation consistency, the orientation of segment $l$ should be opposite respect to $t$ and $t'$. As in Fig.\ref{fig:2d_eg}, $E_l(t)=E_{p_2p_3}(t)$ and $E_l(t')=E_{p_3p_2}(t')$. The first equation in Eq.(\ref{eq:EOEUOU}) implies $E_l(t)=-\Omega_{tt'}E_l(t')$ or $E_{p_2p_3}(t)=\Omega_{tt'}E_{p_2p_3}(t')$. We can also see from Fig.\ref{fig:2d_eg}, the outgoing normals $U(t)$ and $U(t')$ should satisfy $\hat{U}(t)=-\Omega_{tt'}\hat{U}(t')$ such that the basis $\{U(t),E_l(t)\}$ and $\{U(t'),E_{l}(t')\}$ are in the same orientation.

  Next we will prove the following proposition to show that Eq.(\ref{eq:VeV}) is satisfied.
  \begin{Proposition}
    Given two neighboring 4-simplexes $\sigma_v$ and $\sigma_{v'}$, as in Fig.\ref{fig:Glue4simplex}, if the orientation consistency Eq.(\ref{eq:OrientConsist}) and the parallel transportation Eq.(\ref{eq:EOEUOU}) are satisfied, and $\Omega_{vv'}\in\text{SO(4)}$, then $\mathrm{sgn}V_4(v)=\mathrm{sgn}V_4(v')$.
  \end{Proposition}
  \startproof Without losing generality, we assume $\epn^{e_1e_2e_3e_4e_5}(v)=1$. For convenience, we introduce the shorthand notations: $E_{ij}\equiv E_{e_ie_j}(v)$, $E'_{ij}\equiv E_{e'_ie'_j}(v')$, $U^1\equiv U^{e_1}(v)$, $U'^1\equiv U'^{e_1}(v)$. The 4-volumes of $\sigma_v$ and $\sigma_{v'}$ are given by Eq.(\ref{eq:V4})
  \begin{eqnarray*}
    V_4(v)&=&-\det(E_{12},E_{32},E_{42},E_{52})\\
    V_4(v')&=&\det(E'_{12},E'_{32},E'_{42},E'_{52})
  \end{eqnarray*}
where the minus sign for $V_4(v)$ is because of the orientation of $\sigma_v$ is $[e_1,\cdots,e_5]$ while the orientation of $\sigma_{v'}$ is $-[e'_1,\cdots,e'_5]$. By using Eq.(\ref{eq:orthogonal}), we have
  \begin{equation*}
    \frac{1}{3!}U_{I'}^1\epn^{I'JKL}\det(E_{12},E_{32},E_{42},E_{52})=E_{32}^{[J}E_{42}^KE_{52}^{L]}
  \end{equation*}
  Multiply with $\hat{U}^1_I\epn^{IJKL}$ and use $\epn^{IJKL}\epn_{I'JKL}=3!\delta^I_{I'}$, then we have
  \begin{eqnarray}
   &&{\hat{U}_1^I\hat{U}^1_I}{|U^1|}\det(E_{12},E_{32},E_{42},E_{52})\nonumber\\
   &=&\det(\hat{U}_1,E_{32},E_{42},E_{52})\nonumber
  \end{eqnarray}
  Using this result to both $\sigma_v$ and $\sigma_{v'}$, we can easily get
  \begin{eqnarray*}
    \mathrm{sgn}V_4(v)&=&-\sgn\det(\hat{U}_1,E_{32},E_{42},E_{52})\\
    \mathrm{sgn}V_4(v')&=&\sgn\det(\hat{U}'_1,E'_{32},E'_{42},E'_{52})
  \end{eqnarray*}
  By Eq.(\ref{eq:EOEUOU}) and $\det\Omega_{vv'}=1$, we obtain
  \begin{equation*}
    \mathrm{sgn}V_4(v)=\mathrm{sgn}V_4(v')
  \end{equation*}
  $\Box$

  The orientation consistency means if we want to glue two 4-simplexes together, the orientation bivectors in $t_{e_1}$ should be opposite $A_{e_1e_i}(v)=-\Omega_{vv'}\rhd A_{e_1e'_i}(v')$ but the non-oriented bivectors stay the same $A_{e_ie_je_k}(v)=\Omega_{vv'}A_{e'_ie'_je'_k}(v')$. This can be seen from Eq.(\ref{eq:DefA}) and Eq.(\ref{eq:EOEUOU}).

  \subsection{Discrete Geometry of Boundary}
\begin{figure}[htbp!]
    \centering
    \includegraphics[width=0.25\textwidth]{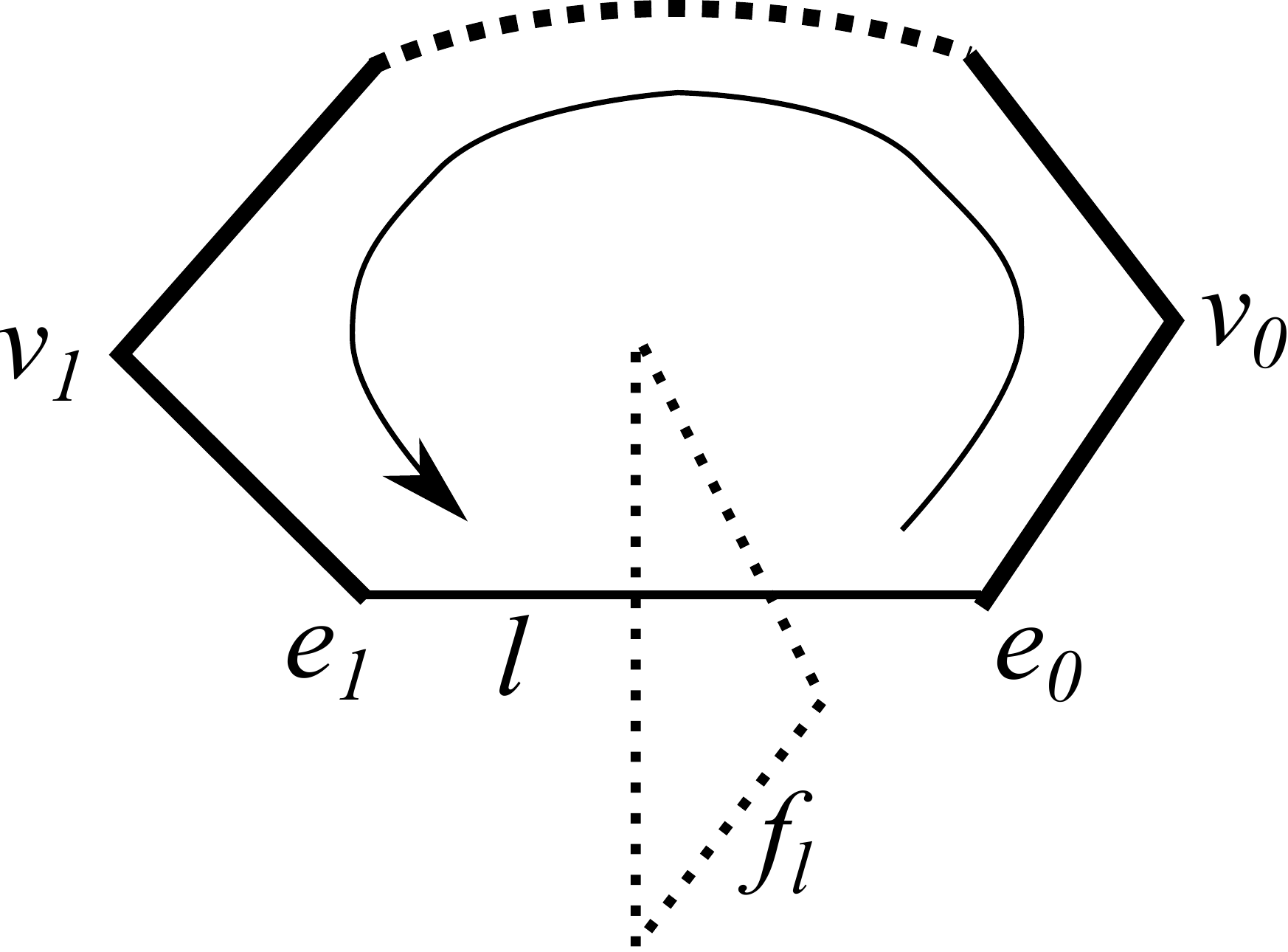}
    \caption{A boundary triangle $f_l$ which is shared by tetrahedron $t_{e_0}$ and $t_{e_1}$ }\label{fig:boundary}
  \end{figure}

  Now we consider the discrete geometry of the boundary of a given simplicial complex $\Delta$. We denote the boundary of $\Delta$ as $\p\Delta$. On $\p\Delta$, each boundary triangle is exactly shared by two boundary tetrahedrons, as shown in Fig.\ref{fig:boundary}.

  At each boundary node $e$ (the center of a boundary tetrahedron $t_e$) we can also construct the segment vectors as before. For each segment $l$ of a boundary tetrahedron $t_e$, we associate it with an oriented vector $E_l(e)$ at $e$. The segment vectors $E_l(e)$ should satisfy the following properties:
  \begin{itemize}
    \item[Inverse:] When the orientation of $l$ is inverted,
    \begin{equation}\label{eq:inverseB}
      E^I_{-l}(e)=-E^I_l(e)
    \end{equation}
    \item[Close:] $\forall f\in\p\Delta$, if its boundary $l_1,~l_2,~l_3$ orientations are consistent, then
    \begin{equation}\label{eq:CloseB}
      E^I_{l_1}(e)+E^I_{l_2}(e)+E^I_{l_3}(e)=0
    \end{equation}
    \item[Gluing:] If edge $e$ touch vertex $v$, $\forall f\in\p t_e,~ l,l' \in\p f$
    \begin{equation}\label{eq:gluingB}
      \delta_{IJ}E_{l}^I(e)E_{l'}^J(e)=\delta_{IJ}E_{l}^I(v)E_{l'}^J(v).
    \end{equation}
    \item[Gauge:] $\forall l\in t_e$, the segment vector $E_l(e)$ is orthogonal to the unit vector $u=(1,0,0,0)$
    \begin{equation}\label{eq:GaugeB}
      E_l(e)\cdot u=0
    \end{equation}
  \end{itemize}
  As before we can also define the induced boundary metric by the boundary segment vectors
  \begin{equation}
    g_{l_1l_2}(e)= \delta_{IJ}E_{l_1}^I(e)E_{l_2}^J(e).
  \end{equation}

  For each boundary tetrahedron $t_e$, it lies in the 3-dimensional subspace which is orthogonal $u$. An oriented tetrahedron $t_e$ can be represented by its ordered four points $[p_1,p_2,p_3,p_4]$. The orientation of $t_e$ should be identified with the induced orientation from the 4-simplex $\sigma_v$ containing $t_e$, i.e. $[p_1,p_2,p_3,p_4]\leftarrow[p_1,p,p_2,p_3,p_4]$, $p\in\sigma_v, p\not\in t_e$. We assume all the tetrahedrons are non-degenerate. Then we can define the oriented 3-volume of $t_e$
  \begin{equation}\label{eq:3V}
    V_3(e)=\frac{1}{3!}\sum_{j,k,l}\epn^{ijkl}(ev)\epn_{JKL}E_{p_jp_i}^J(e)E_{p_kp_i}^K(e)E_{p_lp_i}^L(e)
  \end{equation}
  where $\epn_{JKL}\equiv \epn_{IJKL}u^I$, and $\epn^{ijkl}(e_1v)=\epn^{i1jkl}(v)$.

  Then we can define the 3-vector $n_{p_j}(e)$ which is normal to the face $f\in t_e$ and $f\cap p_j=\emptyset$
  \begin{equation}\label{eq:nfout}
    n^{p_j}_I(e)\equiv\frac{1}{2V_3(e)}\sum_{k,l}\epn^{ijkl}(ev)\epn_{IJK}E_{p_kp_i}^J(e)E_{p_lp_i}^K(e)
  \end{equation}
  which implies
  \begin{equation}
    n^{p_i}_I(e)E_{p_jp_k}^I(e)=\delta^i_j-\delta^i_k
  \end{equation}
It is not hard to show the following relations:
  \begin{eqnarray}
    &&\sum_{i=1}^4n_{p_i}(e)=0\\
    &&V_3(e)^{-1}=\frac{1}{3!}\sum_{j,k,l}\epn_{ijkl}(ev)\epn^{IJK}n^{p_j}_I(e) n^{p_k}_J(e) n^{p_l}_K(e)\label{eq:ntoV3}\\
   && E_{p_ip_j}^I(e)=\frac{V_3(e)}{2}\sum_{k,l}\epn_{ijkl}(ev)\epn^{IJK}n_J^{p_k}(e)n_K^{p_l}(e).\label{eq:ntoE}
  \end{eqnarray}

  For the boundary edge $e$ connecting a vertex $v$, and for any triple of segments $l_1,l_2,l_3 \in t_e$, we have the segment vectors at vertex $v$: $E_{l_i}(v)$ and at $e$: $E_{l_i}(e)$, where $i=1,2,3$. If we consider the unit vector $\hat{U}_e(v)$ defined before, which is orthogonal to $E_{l_i}(v)$ such that
  \begin{equation}\label{uU}
    \begin{split}
      &\text{sgn}~\det\left(E_{l_1}(v),E_{l_2}(v),E_{l_3}(v),\hat{U}_{e}(v)\right)\\     =&\sgn~\det\left(E_{l_1}(e),E_{l_2}(e),E_{l_3}(e),\ep u \right)
    \end{split}
  \end{equation}
  where $\ep=\pm1$. Then there exists a unique SO(4) \cite{Han:2011AsLorentz} matrix $\Omega_{ve}$ such that
  \begin{equation}\label{eq:EOEUOUB}
    (\Omega_{ve})^I_{~J}E^J_l(e)=E^I_l(v),\quad (\Omega_{ve})^I_{~J} u^J=\ep\hat{U}_{e}(v)
  \end{equation}
  Then we find the 3-volume defined in Eq.(\ref{eq:3V}) is consistent with the one induced from $\sigma_v$ up to a sign, i.e. $\ep V_3(v)=V_3(e)$, while the 3-volume of tetrahedron $t_{e_p}$ induced from $\sigma_v$ is defined by
  \begin{equation}\label{eq:V3v}
     V^p_3(v)=\frac{1}{3!}\sum_{j,k,l}(\epn^{ipjkl}\epn_{IJKL}\hat{U}^I_{e_p}E_{e_je_i}^JE_{e_ke_i}^KE_{e_le_i}^L)(v)
  \end{equation}
  
  We can also find an explicit expression between $V_4(v)$ and $V_3(v)$ of $t_e$.
  \begin{equation}
    V_3(v)=V_4(v)\hat{U}_e^I(v)U_I^e(v)
  \end{equation}
  where $\hat{U}^e(v)= U^e(v)/|U^e(v)|$. Because of this $U_I^e(v)$ can also be written as
  \begin{equation}
    U_I^e(v)=\frac{V_3(v)}{V_4(v)}\hat{U}_I^e(v)
  \end{equation}
  Because of $\det\Omega_{ev}=1$ and $\ep\mathrm{sgn}V_3(v)=\mathrm{sgn}V_3(e)$, we have $\mathrm{sgn}(\hat{U}_e^I(v)U_I^e(v))=\ep\mathrm{sgn}(u^IU_I^e(e))$. Further because of $\hat{U}_I^e(v)\hat{U}^I_e(v)=1$, we obtain $\mathrm{sgn}V_3(v)=\mathrm{sgn}V_4(v)$.

The above construction for a boundary tetrahedron can be also extended to any internal tetrahedron $t_e$. As before, we can  construct the segment vectors $E_l(e_1)$ for any segment $l\in t_{e_1}$. However this time for each edge we have two segment vectors $E_l(e_1)$ and $E'_l(e_1)$ associated with $\sigma_v$ and $\sigma_{v'}$ respectively. Because of orientation consistency, $E_l(e_1)=-E'_l(e_1)$ where the minus sign comes from the opposite orientations of $l$ induced from different 4-simplices. Moreover because of Eq.(\ref{eq:EOEUOUB}), we also obtain
  \begin{equation}
    \hat{U}_{e_1}(e_1)=-\hat{U}'_{e_1}(e_1)
  \end{equation}
  where $\hat{U}_{e_1}(e_1)\equiv\Omega_{e_1v}\hat{U}_{e_1}(v)$ and $\hat{U}'_{e_1}(e_1)\equiv\Omega_{e_1v'}\hat{U}_{e_1}(v')$. This relation implies that given two neighboring 4-simplexes share a tetrahedron $t_e$, if $\hat{U}_{e}(e)$ is future-pointing, then $\hat{U}'_{e}(e)$ is past-pointing., or vice versa. From the relation between $E_l(e_1)$ and $E'_l(e_1)$, $\hat{U}_{e_1}(e_1)$ and $\hat{U}'_{e_1}(e_1)$, using Eq.(\ref{eq:V3v}) and $\det\Omega_{ev}=1$, we have
  \begin{equation}
    V_3(e)=V'_3(e).
  \end{equation}

Come back to the boundary tetrahedrons, because of Eq.(\ref{eq:gluing}), Eq.(\ref{eq:gluingB}), for two boundary tetrahedrons $t_{e_0}$ and $t_{e_1}$ that share the triangle $f$, we can get that the induced metric on the triangle $f$ from $t_{e_0}$ and $t_{e_1}$ are the same
  \begin{equation}
    \delta_{IJ}E_{l_1}^I(e_0)E_{l_2}^J(e_0)=\delta_{IJ}E_{l_1}^I(e_1)E_{l_2}^J(e_1)
  \end{equation}
  for any pair of the segments $l_1,l_2$ of the triangle $f$.

  For gluing the boundary tetrahedrons $t_{e_0}$ and $t_{e_1}$, the orientation of the tetrahedrons should be consistent with each other. If the induced orientation of the face $f_l=t_{e_0}\cap t_{e_1}$ are opposite respecting to $t_{e_0}$ and $t_{e_1}$, i.e. $\epn_{pp_1p_2p_3}(e_0)=-\epn_{p'p_1p_2p_3}(e_1)$, the orientations of the two tetrahedrons are consistent.

  For defining the dihedral angle of face $f$, we assign a reference frame at the boundary face $f$. In $f$ frame, we construct the segment vectors $E_l(f)$ for all $l\in f$. Because of the orientation consistency, for each segment $l\in f$, we can define two segment vectors $E_l(f)$ and $E'_l(f)$ respected to two boundary tetrahedrons sharing face $f$. They are opposite since the opposite induced orientations on $l$ from two different tetrahedrons, i.e.
  \begin{equation}
    E_l(f)=-E'_l(f).
  \end{equation}
  Except for satisfying inverse, close, gluing, gauge properties as $E_l(e)$, $E_l(f)$ also satisfies the face gauge, which means there exist a vector $z=(0,0,0,1)$ such that
  \begin{equation}
    E_l(f)\cdot z=0, \quad \forall l\in f
  \end{equation}
  If we consider a normal vector $n_{ef}$ orthogonal to triangle $f$ and $u$ such that
  \begin{equation}
    \begin{split}
      &\text{sgn}~\det\left(E_{l_1}(e),E_{l_2}(e),n_{ef},u\right)\\     =&\text{sgn}~\det\left(E_{l_1}(f),E_{l_2}(f),z,u \right)
    \end{split}
  \end{equation}
  there must be a unique SO(3) matrix $\omega_{ef}$ such that
  \begin{equation}
    \omega_{ef} \rhd E_{l_1}(f)= E_{l_1}(e), \quad \omega_{ef} \rhd z= n_{ef}
  \end{equation}
  Then for the loop holonomy $\Omega_f(f)=\omega_{fe_0}\Omega_{e_0e_1}\omega_{e_1 f}$ we always have
  \begin{equation}
    \Omega_f(f)\rhd E_{p_ip_j}(f)=E_{p_ip_j}(f),\quad \forall l=p_ip_j\in f
  \end{equation}
  where $E_{p_ip_j}(f)\equiv E_l(f)=-E'_l(f)$. Then we know that the non-oriented bivector (the triangle $f=(p_3,p_4,p_4)$)
  \begin{equation}
  A_{e_3e_4e_5}(f)=\frac{1}{2}(E_{e_3e_5}(f)\wedge E_{e_4e_5}(f))
  \end{equation}
  is invariant under the operation of loop spin connection $\Omega_f$, i.e. $A_{e_3e_4e_5}(f)=\Omega_f(f)\rhd A_{e_3e_4e_5}(f)$.

  \subsection{Regge action from connection formalism}
  The construction in the previous subsection is essentially a connection formalism for discrete classical geometry both in the bulk and on the boundary. Here we show how to relate the Regge action from this formalism.

  in order to writing down the Regge action, we should define the deficit angle $\Theta_f$ for internal faces and dihedral angle $\Theta_f^B$ for boundary faces. Let us first consider the internal faces $f$. The first step is to write down the explicit expression for the loop spin connections $\Omega_f$ along the boundary of an internal faces $f$. For an internal face $f$, the loop spin connection keeps the three segment vectors $E_{p_ip_j}(v)$ unchanged by Eq.(\ref{eq:DefA}) and Eq.(\ref{eq:EOEUOU}), where $p_i,p_j$ are the vertices of the triangle $f$,
  \begin{equation}
    \Omega_f(v)E_{p_ip_j}(v)=E_{p_ip_j}(v)
  \end{equation}
  where $\Omega_f(v)\equiv\Omega_{vv'}\cdots \Omega_{v''v}$. The loop spin connection keeps the vectors lying on the plane determined by $E_{p_ip_j}(v)$. It implies that the loop spin connection $\Omega_f(v)\in\text{SO(4)}$ is either a pure boost with a parameter $\theta_f$ or a pure boost connecting $-1\in\text{SO(4)}$ combined with a $\pi$-rotation on the plane determined by $E_{p_ip_j}(v)$, explicitly,
  \begin{equation}
    \Omega_f(v)=e^{in_f\pi}\exp(\theta_f \star \hat{A}_f(v)+n_f\pi\hat{A}_f(v))
  \end{equation}
  where $A_f(v)\equiv A_{e_1e_2e_3}(v)$ is the non-oriented bivector associated to $f$, and $n_f=0,1$. Then we parallel transport $\Omega_f(v)$ it to a neighboring tetrahedron $t_e$ by using $\Omega_{ve}$, i.e. $\Omega_f(e)=\Omega_{ev}\Omega_f(v)\Omega_{ve}$. We find the parameter $\theta_f$ is related to the deficit angle. An explicit way to see it is the following: The curvature in the discrete setting is given by the pure boost part of the above spin connection, i.e. the above $\Omega_f(v)$ with $n_f=0$. Thus to find the relation between the parameter $\theta_f$ with the deficit angle, in the following we only consider $\Omega_f(v)$ with $n_f=0$ which is the pure boost part of the spin connection \cite{Han:2011AsLorentz}. Let $\Omega_f(e)\equiv\Omega_{ee}$ act on the vector $u$ by using Eqs.(\ref{eq:so42su2}) and (\ref{eq:SO4toSU2}), we have
  \begin{equation}
    \begin{split}
      (\Omega_{ee}\rhd u)_I\sigma^I_E&\equiv \Omega_{ee}^-(\Omega_{ee}^+)^{-1}\\
      &=\cos\theta_f\mathds{1}+\sin\theta_f\mathbf{n}_{f}\cdot \mathbf{\sigma}_E
    \end{split}
  \end{equation}
  where $\mathbf{n}_f$ is the unit vector orthogonal to triangle $f$. It is consistent with the orientation of non-oriented area bivector $A_f$. Then we can get
  \begin{equation}
    \cos\Theta_f:=u\cdot \Omega_{ee}\rhd u=\cos\theta_f
  \end{equation}
which implies $\theta_f=\pm\Theta_f$. It is not the case that $\theta_f=\Theta_f$ always holds. Suppose we assume the parameter $\theta_f$ would be a Regge deficit angle being a function of segment lengths only, we make a global parity transformation $E_l(v)\mapsto\mathbf{P}E_l(v)$, and correspondingly for the spin connection $\Omega_f\mapsto\mathbf{P}\Omega_f\mathbf{P}$. Then
  \begin{equation}
    \mathbf{P}\Omega_f\mathbf{P}=\mathbf{P}\exp(\theta_f\star A_f)\mathbf{P}=\exp(-\theta_f\star \mathbf{P}\rhd A_f)
  \end{equation}
implies $\theta_f\mapsto -\theta_f$ under the parity transformation, where the above second equality is because $\mathbf{P} \star=-\star\mathbf{P}$. However, the parity transformation does not change the segment lengths. Therefore we see that $\theta_f$ does not only depend on the segment lengths. In order to give the relation between $\theta_f$ and deficit angle $\Theta_f$, let us see the discrete version of Einstein-Hilbert action $S=\int\dd^4 x\sqrt{g}R/2$. For each dual face $f$
  \begin{equation}
    \begin{split}
      S_f&=\frac{1}{2}\tr\bigg(\int_{\Delta_f}\sgn\det(e)\star(e\wedge e)\int_fR\bigg)\\
      &\simeq \sgn(V_4)\frac{1}{2}\tr\bigg(\star A_f(e)\ln\Omega_{ee}\bigg)=\sgn(V_4)A_f\theta_f
    \end{split}
  \end{equation}
  where $A_f$ is the face area of triangle $f$ and $\sgn(V_4)$ is the sign of the four volume of the simplexes. Recall that Regge action $S_f=A_f\Theta_f$ is the discretization of Einstein-Hilbert action, we find
  \begin{equation}\label{eq:DefDifA}
    \Theta_f=\sgn(V_4)\theta_f
  \end{equation}
$\Theta_f$ is the deficit angle of interior face $f$, which measures the curvature located at the triangle $f$.

  Now let us consider the case of a boundary face $f$. The relation $E_{p_ip_j}(f)=\Omega_f(f)\rhd E_{p_ip_j}(f)$ ($p_j,p_j$ are the vertices of the triangle $f$) implies that $\Omega_f(f)$ can be written in terms of the non-oriented area bivector $A_f(f)\equiv A_{e_3e_4e_5}(f)$ as
  \begin{equation}
    \Omega_f(f)=e^{in_f\pi}\exp(\theta_f^B \star\hat{A}_f(f)+n_f\pi\hat{A}_f(f))
  \end{equation}
with $n_f=0,1$. Only the pure boost part of the $\Omega_f(f)$ contributes the extrinsic curvature on the boundary, so we only consider the case with $n_f=0$ \cite{Han:2011AsLorentz}. Then the spin connection becomes
  \begin{equation}
    \Omega_{e_0e_1}=\omega_{e_0 f}\exp(\theta_f^B \star\hat{A}_f(f))\omega_{f e_1}
  \end{equation}
Acting $\Omega_{e_0e_1}$ on the vector $u=(1,0,0,0)$, we obtain the dihedral angle $\Theta^B_f=\pm\theta_f^B$ by
  \begin{equation}
    \cos\Theta_f^B=u\cdot\Omega_{e_0e_1}\rhd u=\cos\theta_f^B
  \end{equation}
By a similar discussion as we just did for the deficit angle, $\theta_f^B$ is not exactly the dihedral angle defined in Regge calculus since it is changed under parity transformation. The relation between $\theta_f^B$ and dihedral angle $\Theta_f^B$ is given by
  \begin{equation}\label{eq:DefDiA}
    \sgn(V_4)\Theta_f^B=\theta_f^B
  \end{equation}
  A detail discussion about this relation can be found in \cite{Han:2011AsLorentz} (see also \cite{Barrett:2009gg}\cite{Barrett:2009mw}). Here the spin connection is then given by the following dihedral rotation on the plane orthogonal to the triangle $f$
  \begin{equation}
     \Omega_f(f)=\exp(\sgn(V_4)\Theta_f^B \star\hat{A}_f(f)).
  \end{equation}


  Now we give a brief summary of the the section. In this section we worked on a global oriented simplicial complex $\Delta$ and defined discrete geometric variables segment vectors $E_l(v)$, $E_l(e)$ and $E_l(f)$ at each vertex $v$, boundary edge $e$ and boundary face $f$, respectively. They are the natural (co)-frames for the discrete geometry. They all satisfy the properties of inverse (Eqs.(\ref{eq:inverse}), (\ref{eq:inverseB})), close (Eqs.(\ref{eq:Close}) and (\ref{eq:CloseB})) and gluing (Eqs.(\ref{eq:gluing}) and (\ref{eq:gluingB})). There is a discrete metric $g_{ll'}$ defined by $E_l$ and $E_{l'}$ which is the segment length when $l=l'$. We assume the oriented 4-volume $V_4(v)$ has a constant sign $\sgn V_4(v)$ on the entire complex. From $E_l(v)$ we can define five outpointing vectors $U(v)$ for each $\sigma_v$ which satisfy Eqs.(\ref{eq:closeU}) and (\ref{eq:orthogonal}). For two neighboring simplexes $\sigma_v$ and $\sigma_{v'}$, the their frames are related by SO(4) spin connections $\Omega_{vv'}$. The segment vectors $E_l(v)$ and $E_l(v')$, the unit outpointing vectors $\hat{U}(v)$ and $\hat{U}(v')$ are related by parallel transportation Eq.(\ref{eq:EOEUOU}). The deficit angle and dihedral angle are defined from the spin connection by Eqs.(\ref{eq:DefDifA}) and (\ref{eq:DefDiA}) respectively.

  In the following sections we discuss the asymptotic behavior of Euclidean EPRL spin foam amplitude. We will use the critical configurations $\{j_f, n_{ef}, g_{ve}\}$ to construct (semi-)geometrical variables and to compare them with the ones introduced in this section.

  \section{Equations of Motion}\label{sec:EoM}

  As we discussed in Section \ref{sec:SemiCd}, the asymptotic behavior of Euclidean spin foam amplitude is critical configurations that solve Eqs.(\ref{eq:aR0}), (\ref{eq:ag0}), and (\ref{eq:an0}). The presentation in the following is for the case with Barbero-Immirzi parameter $\gamma<1$. However it turns out that the case with $\gamma>1$ results in the same equations of motion thus the same geometric interpretation.

  Firstly, we consider Eq.(\ref{eq:aR0}). Using the definition Eq.(\ref{eq:actionE}), we get
  \begin{equation}
  \begin{split}
    \Rl{S}&=\sum_{f}\sum_{v\in f}\sum_{\pm}2j_{f}^{\pm }\ln \frac{1+R\left( g_{ve}^{\pm}\right) \mathbf{n}_{ef}\cdot R\left( g_{ve^{\prime }}^{\pm }\right) \mathbf{n}_{e^{\prime }f}}{2}\\
    &=0
  \end{split}\nonumber
  \end{equation}
where $R\left( g\right)$ is the vector representation of $g\in\text{SU(2)}$. The above equation results in
  \begin{equation}\label{eq:gluingEv}
    R\left( \bar{g}_{ve}^{\pm }\right) \mathbf{n}_{ef}=R\left( \bar{g}_{ve^{\prime }}^{\pm}\right) \mathbf{n}_{e^{\prime }f}
  \end{equation}
  which is called \emph{gluing condition} between tetrahedrons.

  Secondly, we consider Eq.(\ref{eq:ag0}). Here we parameterize the group element $g^{\pm}$ by Euler angles $\theta_i,~ i=1,2,3$ around the stationary point $\bar{g}^\pm$, i.e. $g^{\pm}=\exp(\theta_i^{\pm}J^i)\bar{g}^{\pm}$. Evaluate the derivatives over $\theta_i$ on the constraint surface of Eq.(\ref{eq:gluingEv}), we get the following closure condition
  \begin{equation}
    \begin{split}
      \frac{\partial S}{\partial \mathbf{\theta }^{ve}}|_{\theta ^{ve}=0}&=\sum_{f_{n}\in e}^{n}j_{f}^{\pm }\frac{\left\langle n_{ef}\left\vert g_{ev}^{\pm }\left( -\frac{1}{2}\mathbf{\sigma }\right) g_{ve^{\prime}}^{\pm }\right\vert n_{e^{\prime }f}\right\rangle }{\left\langle n_{ef}\left\vert g_{ev}^{\pm }g_{ve^{\prime }}^{\pm }\right\vert n_{e^{\prime }f}\right\rangle } \\
      &\quad+\sum_{f_{4-n}\in e}^{4}j_{f}^{\pm }\frac{\left\langle n_{e^{\prime}f}\left\vert g_{e^{\prime }v}^{\pm }\frac{1}{2}\mathbf{\sigma }g_{ve}^{\pm}\right\vert n_{ef}\right\rangle }{\left\langle n_{e^{\prime }f}\left\vert g_{e^{\prime }v}^{\pm }g_{ve}^{\pm }\right\vert n_{ef}\right\rangle } \\
      &=\sum_{f\in e}^{4}2\ep_{ef}(v)j_{f}^{\pm }R\left( g_{ve}^{\pm }\right)\mathbf{n}_{ef}=0
    \end{split}
  \end{equation}
  where $\ep_{ef}(v)=1$ when the orientations of $f$ and $e$ are agree, otherwise $\ep_{ef}(v)=-1$. As we defined in Section \ref{sec:Def}, the orientation of the half-edges are always from $e$ to $v$. It implies
  \begin{equation}\label{eq:epsv}
    \ep_{ef}(v)=-\ep_{ef}(v'), \quad \ep_{ef}(v)=-\ep_{e'f}(v).
  \end{equation}

  Finally we consider Eq.(\ref{eq:an0}). Here we introduce the derivative of the coherent state $|n\rangle$. Since $|n\rangle,|Jn\rangle$ is a basis of the spinor space $\mathbb{C}^2$ and the spinor $|n\rangle$ is normalized, we have
  \begin{eqnarray}\label{eq:dn}
    \delta |n\rangle &=&\varepsilon |Jn\rangle +i\eta |n\rangle \\
    \delta \langle n| &=&\langle Jn|\bar{\varepsilon}-i\eta \langle n|
  \end{eqnarray}
  where the parameters $\varepsilon\in\mathbb{C}$ and $\eta\mathbb{R}$, $J$ is an anti-linear map defined in \cite{Barrett:2009mw}\cite{Roberto:SFM}
  \begin{equation}\label{eq:J}
    |Jn\rangle\equiv J\binom{z_{0}}{z_{1}}=\binom{-\bar{z}_{1}}{\bar{z}_{0}}
  \end{equation}
  From this definition we can find $|Jn\rangle$ is orthogonal to $|n\rangle$ because $\braket{n}{Jn}=0$. Recall Eq.(\ref{eq:defCS}), the map $J$ sends the 3-vector $\mathbf{n}$ to $-\mathbf{n}$.

  Evaluating Eq.(\ref{eq:an0}) with the derivative Eq.(\ref{eq:dn}) while taking Eq.(\ref{eq:gluingEv}) into account, we find Eq.(\ref{eq:an0}) is satisfied automatically
  \begin{equation*}
    \begin{split}
      \delta _{n_{ef}}S &=j_{f}^{\pm }\delta _{n_{ef}}\left( \ln \left\langle n_{e''f}\left\vert g_{e''v}^{\pm }g_{ve}^{\pm }\right\vert n_{ef}\right\rangle\right)\\
      &\quad+j_{f}^{\pm }\delta _{n_{ef}}\left( \ln \left\langle n_{ef}\left\vert g_{ev}^{\pm }g_{ve^{\prime }}^{\pm }\right\vert n_{e^{\prime }f}\right\rangle \right) \\
      &=2j_{f}^{\pm }\frac{\varepsilon \left\langle n_{e''f}\left\vert g_{e''v}^{\pm }g_{ve}^{\pm }\right\vert Jn_{ef}\right\rangle \left\langle n_{ef}\left\vert g_{ev}^{\pm }g_{ve^{\prime }}^{\pm }\right\vert n_{e^{\prime }f}\right\rangle }{\left\langle n_{e''f}\left\vert g_{e''v}^{\pm}g_{ve}^{\pm }\right\vert n_{ef}\right\rangle \left\langle n_{ef}\left\vert g_{ev}^{\pm }g_{ve^{\prime }}^{\pm }\right\vert n_{e^{\prime}f}\right\rangle } \\
      &\quad+2j_{f}^{\pm }\frac{\bar{\varepsilon}\left\langle n_{e''f}\left\vert g_{e''v}^{\pm }g_{ve}^{\pm }\right\vert n_{ef}\right\rangle \left\langle Jn_{ef}\left\vert g_{ev}^{\pm }g_{ve^{\prime }}^{\pm }\right\vert n_{e^{\prime }f}\right\rangle }{\left\langle n_{e''f}\left\vert g_{e''v}^{\pm}g_{ve}^{\pm }\right\vert n_{ef}\right\rangle \left\langle n_{ef}\left\vert
      g_{ev}^{\pm }g_{ve^{\prime }}^{\pm }\right\vert n_{e^{\prime}f}\right\rangle } \\
      &=0
    \end{split}
  \end{equation*}   
  where the third equality is because
  \begin{eqnarray}
    \left\langle n_{e"f}\left\vert g_{e"v}^{\pm }g_{ve}^{\pm }\right\vert Jn_{ef}\right\rangle \left\langle n_{ef}\left\vert g_{ev}^{\pm}g_{ve^{\prime }}^{\pm }\right\vert n_{e^{\prime }f}\right\rangle &=&0 \\
    \left\langle n_{e"f}\left\vert g_{e"v}^{\pm }g_{ve}^{\pm }\right\vert n_{ef}\right\rangle \left\langle Jn_{ef}\left\vert g_{ev}^{\pm}g_{ve^{\prime }}^{\pm }\right\vert n_{e^{\prime }f}\right\rangle &=&0
  \end{eqnarray}
  are satisfied on the constraint surface of Eq.(\ref{eq:gluingEv}).

  Thus we summarize the equations of motion at the end of the subsection.   Gluing condition:
  \begin{equation}\label{eq:gluingEs}
    g_{ve}^{\pm }\rhd \mathbf{n}_{ef} =g_{ve^{\prime}}^{\pm }\rhd \mathbf{n}_{e^{\prime }f}
  \end{equation}
  Closure condition:
  \begin{equation}\label{eq:closeEs}
    \sum_{f\in e}^{4}\ep_{ef}\left( v\right) j^{\pm}_f\left( g_{ev}^{\pm }\rhd \mathbf{n}_{ef}\right)=0
  \end{equation}
  with the orientation condition:
  \begin{equation}\label{eq:orientEs}
    \ep_{ef}(v)=-\ep_{ef}(v'), \quad \ep_{ef}(v)=-\ep_{e'f}(v)
  \end{equation}
The critical configurations $(j_f,g_{ve},n_{ef})$ are the solutions of the above equations.

  \section{Semi-geometrical Variables}\label{sec:SemiGV}
  In this section, we construct bivector variables at each vertex $v$ in terms of spin foam variables $(j_f, g_{ve},n_{ef})$. We call the bivectors constructed in this section the \emph{Semi-geometrical Variables}.

  We identify any bivectors $X_{IJ}\in\mathbb{E}$ with SO(4) Lie algebra element $J^{IJ}\in\la{so}_4$ by using
  \begin{equation}
    X\equiv X_{IJ}J^{IJ}
  \end{equation}
  As we know, $\la{so}_4$ Lie algebra can be decomposed into two copies $\la{su}_2$ Lie algebra, i.e. self-dual and anti-self-dual parts. Give any $J^{IJ}\in\la{so}_4$ and define $J^i=\frac{1}{2}\epn^{ijk}J^{jk}$, $K^i=J^{i0}$, the generators $J^{\pm i}=\frac{1}{2}(J^i\pm K^i)$ satisfy the following commutation relations
  \begin{equation*}
    \begin{split}
      [J^{\pm i},J^{\pm j}] &= -\epn^{ijk}J^{\pm k}\\
      [J^{\pm i},J^{\mp j}] &= 0
    \end{split}
  \end{equation*}
  The explicit relation between SO(4) (or Spin(4)) group element and SU(2)$\otimes$SU(2) is
  \begin{equation}\label{eq:so42su2}
    \begin{split}
      \exp\left(\frac{1}{2}B_{IJ}J^{IJ}\right)&=\exp\left(\sum_{\pm}(\frac{1}{2}\epn_i{}^{jk}B_{jk}\pm B_{i0})J^{\pm i}\right)\\
      &\equiv\exp\left(\sum_{\pm}B_i^{\pm}J^{\pm i}\right)\\
      \exp\left(\frac{1}{2}(\star B)_{IJ}J^{IJ}\right)&=\exp\left(\sum_{\pm}\mp(\frac{1}{2}\epn_i{}^{jk}B_{jk}\pm B_{i0})J^{\pm i}\right)\\
      &\equiv\exp\left(\sum_{\pm}\mp B_i^{\pm}J^{\pm i}\right)\\
    \end{split}
  \end{equation}

  Based on this decomposition, we define the self-dual and anti-self-dual bivectors in each tetrahedron $t_e$ associated to the faces $f$ of the tetrahedron. The canonical quantization of LQG suggests that the area spectral is given by $\gamma j l_p^2$ when $j$ is much larger than $1$. So here we define the self-dual bivector $X^+_{ef}(e)$ and anti-self-dual bivector $X^-_{ef}(e)$ for the face $f$ in tetrahedron $t_e$ as
  \begin{equation}\label{eq:Xsa}
    \im X^{\pm}_{ef}\equiv 2\gamma j_f(n_{ef})_iJ^{\pm i}=\im(X_{ef})_i\sigma^{\pm i}
  \end{equation}
  Using the above definition we can define the unit bivectors as
  \begin{equation}\label{eq:PTsa}
    \hat{X}^{\pm}_{ef}\equiv\frac{X^{\pm}_{ef}}{|X^{\pm}_{ef}|}= (n_{ef})_i\sigma^i
  \end{equation}
  where $|X|^2\equiv X_iX^i$, $|X^{\pm}_{ef}|=\gamma j_f$. The parallel transportations of $\hat{X}^{\pm}_{ef}$ are
  \begin{equation}\label{eq:PTX}
    \hat{X}^{\pm}_{ef}(v)=g^{\pm}_{ve}\hat{X}^{\pm}_{ef}g^{\pm}_{ev}
  \end{equation}
  Then by using Eq.(\ref{eq:so42su2}) and Eq.(\ref{eq:Xsa}) we can write the SO(4) bivector
  \begin{equation}\label{eq:so4bivector}
    \begin{split}
      X_{ef}&\equiv \sum_{\pm}\im X^{\pm}_{ef}=\gamma j_f\epn_{0IJK}n_{ef}^KJ^{IJ}\\
      &\equiv(X_{ef})_{IJ}J^{IJ}
    \end{split}
  \end{equation}
  where $n_{ef}^K\equiv(0,n_{ef}^k)$. Then we can define the unit bivector
  \begin{equation}
    \hat{X}_{ef}\equiv\frac{X_{ef}}{|X_{ef}|}=\epn_{0IJK}n_{ef}^KJ^{IJ}
  \end{equation}
  where $|X|^2=\frac{1}{2}X_{IJ}X^{IJ}$, $|X_{ef}|=\gamma j_f$. Based on Eq.(\ref{eq:PTsa}), we can parallel transport $X_{ef}$ to the nearest vertex $v$ and define a bivector at $v$ by
  \begin{equation}\label{eq:PTev}
    X_{ef}(v)=g_{ve}\rhd X_{ef}
  \end{equation}

In $t_e$ frame there is a unit vector $u=(1,0,0,0)$ such that
  \begin{equation}\label{eq:SimE}
    \delta^{IJ}u_I(X_{ef})_{JK}=0
  \end{equation}
Any vector $x_I\in\mathbb{E}$ can be identified with a $2\times2$ matrix $x=x_I\sigma_E^I$, where $\sigma_E=(\mathds{1},\im\sigma^i)$. The parallel transformation for this vector is
  \begin{equation}\label{eq:SO4toSU2}
    g^-x_I\sigma_E^I(g^+)^{-1}=(gx)_I\sigma_E^I, \quad \forall g=(g^+,g^-)\in \mathrm{SO(4)}
  \end{equation}
  Then parallel transportation Eq.(\ref{eq:SimE}) to vertex $v$ by $g_{ve}\in$SO(4). By defining $N^e(v)=g_{ve}\rhd u$, we can have $\delta^{IJ}N^e(v)_I(X_{ef}(v))_{JK}=0$ which is the simplicity constraint for each faces at each vertex.

  Now we rewrite equations of motion Eq.(\ref{eq:gluingEs}) and Eq.(\ref{eq:closeEs}) by using SO(4) bivectors $X_{ef}(v)$ and summarize them in the follows:

  Gluing condition:
  \begin{equation}\label{eq:gluingEX}
    X_f(v)\equiv X_{ef}(v)= X_{e^{\prime }f}(v).
  \end{equation}
  Closure condition:
  \begin{equation}\label{eq:closeEX}
    \sum_{f\in e}^{4}\ep_{ef}(v)X_{ef}(v)=0.
  \end{equation}
We can also get two more equations from the definitions.
  In terms of Eq.(\ref{eq:PTev}) we obtain
  \begin{equation}\label{eq:PTvv}
    g_{v^{\prime}v}\rhd X_{ef}(v)=X_{ef}(v^{\prime}).
  \end{equation}
We also have the simplicity constraint
  \begin{equation}\label{eq:SimCstE}
    \delta^{IJ}N^e(v)_I(X_{ef}(v))_{JK}=0.
  \end{equation}

  \section{Discrete Geometry from Critical Configurations}\label{subsec:DGdeSS}
  In this section we use the semi-geometrical variables $X_{ef}(v)$ and $N^{e}(v)$ to reconstruct the discrete geometrical variables $E_{e_ie_j}(v)$ and $U^e(v)$. Here in this section we only discuss the case that all the 4-simplices are non-degenerate (The degenerate case is discussed in Section\ref{sec:AsympD}). In our definition and the definition in \cite{Conrady:SL2008}, the non-degeneracy is defined in terms of $N^e(v)$ by
  \begin{equation}\label{eq:nondeg}
    \prod_{1\leq i< j< k< l}^5 \det(N^{e_i}(v),N^{e_j}(v),N^{e_k}(v),N^{e_l}(v))\neq0
  \end{equation}
The reconstruction of the non-degenerate geometry in the case of a simplicial manifold without boundary was first introduced in \cite{Conrady:SL2008}.

$N^e(v)$ is determined by the group element $g_{ve}$ for a set of given configuration $\{j_f,g_{ve},n_{ef}\}$. For Euclidean theory, as discussed in \cite{Barrett:2009gg}, if Eq.(\ref{eq:nondeg}) is satisfied, $g_{ve}^{+}$ and $g_{ve}^-$ should be two different SU(2) group elements, i.e. $g_{ve}^{+}\neq g_{ve}^-$.

During the following construction, we keep in mind that we are working on an consistently oriented complex $\Delta$, where the orientations of the 4-simplices is defined in Section \ref{sec:DcG}.

  \subsection{Reconstruction of 4-simplex}
The following analysis is based on a given non-degenerate critical configuration $\{j_f, n_{ef}, g_{ve}\}$. In the frame of $v$, we consider two bivectors $X_{ef}(v)$ and $X_{e'f}(v)$. Because of the simplicity constraint Eq.(\ref{eq:SimCstE}), there are 4-D vectors $V_{ef}$ and $V_{e'f}$ in $\mathbb{E}$ such that $\star X_{ef}(v)=N^{e}(v)\wedge V_{ef}(v)$ and $\star X_{e'f}(v)=N^{e'}(v)\wedge V_{e'f}(v)$. Because of the Gluing condition Eq.(\ref{eq:gluingEX}), vectors $N^{e}(v)$, $V_{ef}(v)$, $N^{e'}(v)$ and $V_{e'f}(v)$ are in the same plane. Then this plane is spanned by $N^{e}(v)$ and $N^{e'}(v)$ i.e. $V_{ef}(v)=\alpha_{ee'}(v)N^{e'}(v)+cN^e(v)$. So the bivector $X_f(v)\equiv X_{ef}(v)=X_{e'f}(v)$ can be written as
  \begin{equation}\label{eq:Xf}
    X_f(v)=\star\alpha_{ee'}(v)\left(N^{e}(v)\wedge N^{e'}(v)\right)
  \end{equation}
  In one 4-simplex $\sigma_v$ we can denote $X_f(v)$ by two edges. If a triangle $f$ is shared by two tetrahedron $t_e$ and $t_e'$, we can denote the bivector $X_f$ as $X_{ee'}=X_{e'e}$. We denote the edges that are attaching at $v$ by $e_i$, $i=1,2,3,4,5$. The vertex is oriented as $[e_1,e_2,\cdots,e_5]$. Then Eq.(\ref{eq:Xf}) can be written as
  \begin{equation}\label{eq:XeeE}
    X_{e_ie_j}(v)=\star\alpha_{ij}(v)\left(N^{i}(v)\wedge N^{j}(v)\right)
  \end{equation}

  Using Closure constraint Eq.(\ref{eq:closeEX}), the above equation turns into
  \begin{equation}
    \begin{split}
      \sum_{j\neq i}\ep_{e_ie_j}(v)X_{e_ie_j}(v)&=\star\bigg(N^{i}(v)\wedge \sum_{j\neq i}\ep_{e_ie_j}(v)\alpha_{ij}(v)N^{j}(v)\bigg)\\
      &\equiv\star\bigg(N^{i}(v)\wedge \sum_{j\neq i}\beta_{ij}(v)N^{j}(v)\bigg)\\
      &=0
    \end{split}
  \end{equation}
  where $\ep_{e_ie_j}(v)$ are the coefficients in orientation condition Eq.(\ref{eq:orientEs}) such that $\ep_{e_ie_j}(v)=-\ep_{e_je_i}(v)\equiv\ep_{e_if}(v)=-\ep_{e_jf}(v)$, and  $\beta_{ij}(v)\equiv\ep_{e_ie_j}(v)\alpha_{ij}(v)$. Together with the non-degenerate assumption Eq.(\ref{eq:nondeg}), it implies that there are non-vanishing diagonal elements $\beta_{ii}(v)$ such that
  \begin{equation}\label{eq:closebijN}
    \sum_{j=1}^5\beta_{ij}(v)N^{e_j}(v)=0
  \end{equation}
  Otherwise any two of $N^{e_i}(v)$ would be parallel to each other.

  Now we consider
  \begin{equation}
    \begin{split}
      &\quad\beta _{km}(v)\sum_{j=1}^{5}\beta _{lj}(v)N^{e_{j}}\left( v\right)-\beta_{lm}(v)\sum_{j=1}^{5}\beta _{kj}(v)N^{e_{j}}\left( v\right)  \\
      &=\sum_{j\neq m}\left( \beta _{km}(v)\beta _{lj}(v)-\beta _{lm}(v)\beta _{kj}(v)\right)N^{e_{j}}\left( v\right)=0
    \end{split}
  \end{equation}
  Because of Eq.(\ref{eq:nondeg}), any four $N^{e_i}(v)$ are linearly independent. The above equation turns into
  \begin{equation}
     \beta _{km}(v)\beta _{lj}(v)-\beta _{lm}(v)\beta _{kj}(v)=0
  \end{equation}
  We can pick one $j_0$ for one $\sigma_v$ and ask $l=j=j_0$. Then we can get
  \begin{equation}
    \beta _{km}(v)=\frac{\beta _{m j_0}(v)\beta _{kj_0}(v)}{\beta _{j_0j_0}(v)}\equiv\tilde{\ep}(v)\beta _{m}(v)\beta _{k}(v)
  \end{equation}
  where $\beta _{i}(v)\equiv\beta _{ij_0}(v)/\sqrt{|\beta _{j_0j_0}(v)|}$, $\tilde{\ep}(v)\equiv\mathrm{sgn}(\beta _{j_0j_0}(v))$. Then bivector $\ep_{e_ie_j}(v)X_{e_ie_j}(v)$ becomes
  \begin{equation}\label{eq:XNNE}
    \ep_{e_ie_j}(v)X_{e_ie_j}(v)=\tilde{\ep}(v)\star\left[(\beta_{i}(v)N^{i}(v))\wedge (\beta_{j}(v)N^{j}(v))\right].
  \end{equation}

  Now we can reconstruct the discrete geometrical variable $U^e(v)$ (up to a overall sign in each $\sigma_v$) defined in Section \ref{sec:DcG} by
  \begin{equation}\label{eq:NtoU}
    U^{e_i}(v)\equiv\pm\frac{\sqrt{2}\beta_i(v)N^{e_i}(v)}{\sqrt{|V_4(v)|}}
  \end{equation}
  where $V_4(v)$ is defined by
  \begin{widetext}
    \begin{eqnarray}\label{eq:defV4}
     \frac{V_4(v)}{4}:=\det(\beta_2(v)N^{e_2}(v),\beta_3(v)N^{e_3}(v),\beta_4(v)N^{e_4}(v),\beta_5(v)N^{e_5}(v))
    \end{eqnarray}
  \end{widetext}
from which we obtain
  \begin{equation}
    V_4^{-1}(v)=\det(U^{e_2},U^{e_3},U^{e_4},U^{e_5})
  \end{equation}
  By using Eq.(\ref{eq:UtoE}), we can define segment vectors $E_{e_ie_j}(v)$ satisfying the inverse and close properties, such that Eq.(\ref{eq:XNNE}) turns into
  \begin{equation}\label{eq:RconstAfX}
    \begin{split}
      \ep_{e_ie_j}(v)X_{e_ie_j}(v)&=\tilde{\ep}(v)\frac{1}{2}\left|V_4(v)\right|\star(U^{e_i}(v)\wedge U^{e_j}(v))\\
      &\equiv \ep(v)\frac{1}{2}V_4(v)\star(U^{e_i}(v)\wedge U^{e_j}(v))\\
      &\equiv \ep(v)\frac{1}{4}\sum_{m,n}\epn^{kijmn}E_{e_me_k}(v)\wedge E_{e_ne_k}(v)\\
      &=\ep(v)A_{e_ie_j}\\
      &=\ep(v)\epn_{e_me_ne_ke_ie_j}(v)A_{e_me_ne_k}(v)
    \end{split}
  \end{equation}
  where $\ep(v)\equiv\tilde{\ep}(v)\mathrm{sgn}(V_4(v))$. In the last equality, we use Eq.(\ref{eq:XUtoXE}). This is the explicit relation between semi-geometrical bivector $X(v)$ and discrete geometrical bivectors $A(v)$ defined by Eq.(\ref{eq:DefA}).

  The above result shows that give a set of non-degenerate critical configurations $\{j_f, n_{ef}, g_{ve}\}$ at a vertex $v$, we can reconstruct a bivector geometry in each 4-simplex $\sigma_v$ \cite{Barrett:2009gg}\cite{Barrett:2009mw}.

  \subsection{Gluing the interior 4-simplexes}
In order to construct a discrete geometry on the entire complex, we discuss the gluing of the geometries of two neighboring vertices $v$ and $v'$ that are linked by edge $e_1$. We still use Fig.\ref{fig:Glue4simplex} in our discussion.

  For convenience, we introduce shorthand notations: $U^i\equiv U^{e_i}(v)$, $E_{ij}\equiv E_{e_ie_j}(v)$, $U'^i\equiv g_{vv'}U'^{e'_i}(v')$, $E'_{ij}\equiv g_{vv'}E_{e'_ie'_j}(v')$, with $e'_1\equiv e_1$.

  Because $N^{e_1}(v)=g_{ve}u$, $N^{e_1}(v')=g_{v'e}u$, we have $N^{e_1}(v)=g_{vv'}N^{e_1}(v')$. Together with Eq.(\ref{eq:NtoU}), we have
  \begin{equation}\label{eq:UU}
    \frac{U'^1}{|U'^1|}=\tilde{\ep}\frac{U^1}{|U^1|}
  \end{equation}
  where $\tilde{\ep}=\pm1$. Moreover because of Eqs.(\ref{eq:PTvv}), (\ref{eq:orientEs}), and (\ref{eq:RconstAfX}), we have
  \begin{equation}\label{eq:AA}
    \ep_{e_1e_i}(v)X_{e_1e_i}(v)=\frac{1}{2}\ep V\star(U^1\wedge U^i)=-\frac{1}{2}\ep' V'\star(U'^1\wedge U'^i)
  \end{equation}
  where $i\neq 1$ and we use the shorthand notations $\ep\equiv\ep(v)$, $\ep'\equiv\ep(v')$. Reminding the orientation consistency that is discussed in Section\ref{sec:DcG}, in this notation we should have $\epn_{e_1e_2e_3e_4e_5}(v)=-\epn_{e_1e'_2e'_3e'_4e'_5}(v')$. Then
  \begin{eqnarray}
    V^{-1}&=&\bar{\ep}\det(U^2,U^3,U^4,U^5)\\
    V'^{-1}&=&\bar{\ep}'\det(U'^2,U'^3,U'^4,U'^5)
  \end{eqnarray}
  where $\bar{\ep}=-\bar{\ep}'=\pm 1$. Eq.(\ref{eq:UU}) and Eq.(\ref{eq:AA}) imply that $U'^1$ is proportional to $U^1$ and $U'^i$ are the linear combination of $U^1$ and $U^i$. Explicitly,
  \begin{equation}
    U'^i=-\ep\ep'\tilde{\ep}\frac{|U^1|V}{|U'^1|V'}U^i+a_iU^1
  \end{equation}
  where $a_i$ are the coefficients such that $\sum_{i=1}^5U^i=0$. Using the above equation, we have
  \begin{equation}\label{eq:signs}
    \begin{split}
      \frac{1}{V'}&=\bar{\ep}'\det(U'^2,U'^3,U'^4,U'^5)=\bar{\ep}'\det(U'^1,U'^2,U'^3,U'^4)\\
      &=\tilde{\ep}\frac{U^1}{|U^1|}\left(-\ep\ep'\tilde{\ep}\frac{|U^1|V}{|U'^1|V'}\right)^3\bar{\ep}'\det(U^1,U^2,U^3,U^4)\\
      &=\ep\ep'\left(\frac{|U^1|V}{|U'^1|V'}\right)^2\frac{1}{V'}
    \end{split}
  \end{equation}

  This implies $\ep(v)=\ep(v')$. It means the sign factor $\ep(v)$ defined in Eq.(\ref{eq:RconstAfX}) does not  depend on $v$.
  \begin{Proposition}\label{pro:area}
    For any semi-geometrical area bivector $X_f(v)$ defined in Section \ref{sec:SemiGV} from spin foam model, we can always reconstruct a non-oriented bivector $A_f(v)$ of discrete geometry up to a global sign $\ep$ for the whole simplicial complex $\Delta$.
  \end{Proposition}
  \startproof We prove the proposition in three steps. We can easily seen from Eq.(\ref{eq:RconstAfX}), a semi-geometrical area bivector $X_f(v)$ corresponds to a non-oriented bivector $A_f(v)\equiv A_{e_me_ne_k}(v)$ with a sign factor (recall that $A_{e_me_ne_k}(v)$ is defined with the orientation $f=[e_m,e_n,e_k]$)
  \begin{equation}
    X_{e_ie_j}(v)=\ep(v)\ep_{e_ie_j}(v)\epn_{e_me_ne_ke_ie_j}(v)A_{e_me_ne_k}(v)
  \end{equation}
$\ep_{e_ie_j}(v)\epn_{e_me_ne_ke_ie_j}(v)$ is a sign factor with a given triangle $f=[p_m,p_n,p_k]=[e_m,e_n,e_k]$.  Because $\ep(v)=\ep$ is a global sign, then we focus on proving that  $\ep_{e_ie_j}(v)\epn_{e_me_ne_ke_ie_j}(v)$ is a global sign.

  First we prove that it is a constant for each triangle $f$. For this purpose, we only need to prove the sign is a constant both between two tetrahedrons in one 4-simplex and between two neighboring 4-simplexes, sharing the triangle $f$. We consider the situation showing in Fig.\ref{fig:Glue4simplex}. Because $X_{e_ie_1}(v)=X_{e_1e_i}(v)$, we can get between two tetrahedrons $t_{e_i}$ and $t_{e_1}$ in $\sigma_v$, the sign factor keeps invariant
  \begin{equation}\label{eq:glue2t}
    \ep_{e_ie_1}(v)\epn_{e_me_ne_ke_ie_1}(v)=\ep_{e_1e_i}(v)\epn_{e_me_ne_ke_1e_i}(v)
  \end{equation}
  Because $X_{e_ie_1}(v)=g_{vv'}X_{e'_ie_1}(v')$, $A_{e_me_ne_k}(v)=g_{vv'}A_{e'_me'_ne'_k}(v')$, we can get between two neighboring 4-simplexes $\sigma_v$ and $\sigma_{v'}$, the sign factor keeps invariant
  \begin{equation}\label{eq:glue2s}
    \ep_{e_1e_i}(v)\epn_{e_1e_ie_je_ke_l}(v)=\ep_{e_1e'_i}(v)\epn_{e_1e'_ie'_je'_ke'_l}(v')
  \end{equation}
  Combining Eq.(\ref{eq:glue2t}) and Eq.(\ref{eq:glue2s}), we get for each faces $f$, the sign factor $\ep_{e_ie_j}(v)\epn_{e_me_ne_ke_ie_j}(v)$ is a constant $\ep_f$, once we fix an orientation $[e_m,e_n,e_k]$ of $f$.

  Secondly we prove that the sign factor is a constant in a 4-simplex between different triangles. For this purpose, we only need to prove that for any two triangles in a tetrahedron of a 4-simplex, the sign factor is a constant. We consider the situation showing in Fig.\ref{fig:4simplex}. Without losing generality, we pick out $t_{e_1}\in\sigma_v$ and the bivectors $X_{e_1e_2}(v)$ , $X_{e_1e_3}(v)$. For $X_{e_1e_2}(v)$ we have
  \begin{equation}
    X_{e_1e_2}(v)=\ep(v)\ep_{e_1e_2}(v)\epn_{e_3e_4e_5e_1e_2}(v)A_{e_3e_4e_5}(v)
  \end{equation}
 For $X_{e_1e_3}(v)$ we have
  \begin{equation}
    X_{e_1e_3}(v)=\ep(v)\ep_{e_1e_3}(v)\epn_{e_2e_4e_5e_1e_3}(v)A_{e_2e_4e_5}(v)
  \end{equation}
  If the face orientations of $f_{e_1e_2}$ and $f_{e_1e_3}$ agree along $e_1$,  $\ep_{e_1e_2}(v)=\ep_{e_1e_3}(v)$, We can pick $A_{e_3e_4e_5}(v)$ and $A_{e_5 e_4 e_2}(v)$ (instead of $A_{e_2e_4e_5}(v)$) as the reconstructed non-oriented area bivectors. Then we have
  \begin{equation}
    \ep_{e_1e_2}(v)\epn_{e_3e_4e_5e_1e_2}(v)=\ep_{e_1e_3}(v)\epn_{e_5e_4e_2e_1e_3}(v)
  \end{equation}
  If the face orientations of $f_{e_1e_2}$ and $f_{e_1e_3}$ are opposite at $t_{e_1}$,  $\ep_{e_1e_2}(v)=-\ep_{e_1e_3}(v)$. We can pick $A_{e_3e_4e_5}(v)$ and $A_{e_2 e_4 e_5}(v)$ as the reconstructed non-oriented area bivectors. Then we have
  \begin{equation}
    \ep_{e_1e_2}(v)\epn_{e_3e_4e_5e_1e_2}(v)=\ep_{e_1e_3}(v)\epn_{e_2e_4e_5e_1e_3}(v)
  \end{equation}
  The choice of non-oriented area bivectors can always be achieved based on the orientation of $\Delta^*$. Then in one 4-simplex, the sign factor $\ep_{e_ie_j}(v)\epn_{e_me_ne_ke_ie_j}(v)$ is also a constant $\ep_v=\ep_f$. Then we obtain the following conclusion: for any semi-geometrical bivector $X_f(v)$ constructed from spin foam critical configuration, we can reconstruct a non-oriented bivector of discrete geometry $A_f(v)$, with a choice of the orientation for each $f$, up to a global sign $\ep$ on the entire simplicial complex
  \begin{equation}\label{eq:RconXA}
    X_f(v)=\ep A_f(v)
  \end{equation}
  where $\ep\equiv\ep(v)\ep_f$. $\Box$

  From Eq.(\ref{eq:signs}) we obtain $|U^1|V=\pm|U'^1|V'$. We define a new type of sign factor
  \begin{equation}
  \mu\equiv-\tilde{\ep}|U^1|V/|U'^1|V'=-\tilde{\ep}\mathrm{sgn}(VV').
  \end{equation}
 Recall Eq.(\ref{eq:UtoE}), we obtain
  \begin{equation}
    \begin{split}
      E'^I_{kj}&=V'\epn'_{jklm1}\epn^{IJKL}U'^l_JU'^m_KU'^1_L\\
      &=\mu E^I_{kj}
    \end{split}
  \end{equation}
 which implies that the spin foam variables $g_{vv'}$ and SO(4) holonomy $\Omega_{vv'}$ are just different by a sign
  \begin{equation}\label{eq:RconHL}
    g_{vv'}=\mu_e\Omega_{vv'}
  \end{equation}
  By the definition of the spin connection in Section\ref{sec:DcG}, $\Omega_{vv'}$ is a spin connection as long as $\sgn{V_4(v)}=\sgn{V_4(v')}$.

  \subsection{Reconstruction of boundary}
  First of all, we can reconstruct the tetrahedron $t_e$ with an edge $e$ connecting to the boundary. Giving a set of non-degenerate boundary data $\{j_f, n_{ef}\}$ where $f$s are boundary triangles, we have closure condition for boundary tetrahedron
  \begin{equation}
    \sum_{f\in e}\ep_{ef}(v)\gamma j_f n_{ef}=0
  \end{equation}
  where the $v$ is the vertex $e$ connecting, and $n_{ef}=(0,\mathbf{n}_{ef})$ is lying on the plane orthogonal to $u=(1,0,0,0)$. Then we can reconstruct the discrete geometrical variable $n^p(e)$ defined in Section \ref{sec:DcG} as
  \begin{equation}
    n^{p_i}(e)\equiv \frac{2\ep_{ef_i}(v)\gamma j_{f_i} n_{ef_i}}{|V_3(e)|}
  \end{equation}
  where the oriented 3-volume $V_3(v)$ is defined as
  \begin{widetext}
    \begin{equation}
      \sgn V_3(e) \frac{V^2_3(e)}{8}\equiv\frac{1}{3!}\sum_{j,k,l}\epn_{ijkl}(ev)\epn^{JKL}(\ep_{ef_j}(v)\gamma j_{f_j} n_{ef_j})(\ep_{ef_k}(v)\gamma j_{f_k} n_{ef_k})(\ep_{ef_l}(v)\gamma j_{f_l} n_{ef_l})
    \end{equation}
  \end{widetext}
  In this definition $n^{p_i}(e)$ satisfies Eq.(\ref{eq:ntoV3}).

  Together with Eq.(\ref{eq:ntoE}), we can reconstruct the segment vectors $E_{p_ip_j}(e)$ defined before. Then we can reconstruct the area bivectors Eq.(\ref{eq:so4bivector}) $X_{ef}=\gamma j_f \star(u\wedge n_{ef})$ as
  \begin{equation}
    \begin{split}
      \ep_{ef_i}(v)X_{ef_i}&=\bar{\ep}(e)\frac{1}{2}V_3(e)\star(u\wedge n^{p_i}(e))\\
      &=\bar{\ep}(e)\frac{1}{4}\sum_{j,k}\epn^{lijk}(E_{p_jp_l}(e)\wedge E_{p_kp_l}(e))\\
      &=\bar{\ep}(e)\epn_{e_je_ke_le_i}(ev)A_{e_je_ke_l}(e)
    \end{split}
  \end{equation}
  where $\bar{\ep}(e)=\sgn V_3(e)$, and here all the boundary segment vectors $E_{p_ip_j}(e)$ are understood as 4-vectors in $\mathbb{E}$ such that $E_{p_ip_j}(e)=(0,E_{p_ip_j}^k)$, $k=1,2,3$.

  Now we identify the boundary tetrahedron $t_e$ with the tetrahedron in 4-simplex $\sigma_v$ dual to edge $e$. For convenience, we introduce shorthand notations: $E_{ij}\equiv E_{p_ip_j}(e)$, $E'_{ij}\equiv g_{ev}E_{p_ip_j}(v)$, $\bar{\ep}\equiv\bar{\ep}(e)$, $n^j\equiv n^{p_j}(e)$. Because the parallel transportation $X_{ef}=g_{ev}\rhd X_{ef}(v)$, we have
  \begin{equation}\label{eq:XeXv}
    \bar{\ep}\sum_{j,k}\epn^{lpijk}(E_{jl}\wedge E_{kl})=\ep(v)\sum_{j,k}\epn^{lpijk}(E'_{jl}\wedge E'_{kl})
  \end{equation}
  where $p$ is the point belong to $\sigma_v$ but not in $t_e$. It means
  \begin{equation}
    \bar{\ep}V_3n^i=\ep V'_3n'^i
  \end{equation}
  where $V'_3$ is defined in the same way as $V_3$ but using $E'_{ij}$ instead. Because of $n^iE_{jk}=\delta^i_j-\delta^i_k= n'^iE'_{jk}$, the above equation turns into
  \begin{equation}
    \bar{\ep}V_3n^iE'_{jk}=\ep(v) V'_3n'^iE'_{jk}=\ep(v) V'_3n^iE_{jk}
  \end{equation}
which implies
  \begin{equation}\label{eq:EEprime}
    E'_{jk}=\bar{\ep}\ep(v)\frac{V'_3}{V_3}E_{jk}
  \end{equation}
  Bring this equation back to Eq.(\ref{eq:XeXv}), we get
  \begin{equation}
    \ep(v)\bar{\ep}\left(\frac{V'_3}{V_3}\right)^2=1
  \end{equation}
which implies
  \begin{equation}
    \bar{\ep}(e)=\ep(v)\ \ \ \text{and}\ \ \ |V_3|=|V'_3|
  \end{equation}
$\bar{\ep}(e)=\sgn V_3(e)$ is determined by the boundary data. If we choose the orientations of the triangles such that $\ep_{e_ie_j}(v)\epsilon_{e_me_ne_ke_ie_j}(v)=1$ identically, the global sign $\ep$ relating $X_f(v)$ and $A_f(v)$ is determined by the boundary data $\ep=\sgn V_3(e)$, once we choose $\sgn V_3(e)$ to be a constant on the boundary.

Moreover from Eq.(\ref{eq:EEprime}), we find the spin connection equals the on-shell $g_{ve}$ up to a sign $\mu_e$
  \begin{equation}\label{eq:RconHLB}
    g_{ve}=\mu_e \Omega_{ve}
  \end{equation}
  where $\mu_e=\mathrm{sgn}(V_3)\mathrm{sgn}(V'_3)=\pm 1$. We denote by $V_3^e(v)$ the 3-volume induced from $V_4(v)$ with the normal $\hat{U}_e(v)$. $V_3^e(v)$ is in general different from $V'_3$ by the discussion at Eq.(\ref{uU}). As we show in Section\ref{sec:DcG}, $\sgn V_3^e=\sgn V_4(v)$, we have
  \begin{equation}
    \mu_e=\sgn V_3\sgn V_4(v)\sgn(\hat{U}_e \cdot u).
  \end{equation}
Then we can prove the following lemma,
  \begin{Lemma}
    Given face $f$ either an internal face or a boundary face, the product $\prod_{e\in f}\mu_e$ is invariant when $U_e(v)$ flips sign for any 4-simplex $\sigma_v$, recalling that the five normals $U_e(v)$ have an overall sign ambiguity when reconstructing a 4-simplex $\sigma_v$. Therefore $\prod_{e\in f}\mu_e$ is determined by the spin foam critical configurations.
  \end{Lemma}
  \startproof For an internal edge $e=(vv')$, we have
  \begin{equation}
    \begin{split}
      \mu_e&=-\tilde{\ep}_e\sgn V_4(v)\sgn V_4(v')\\
      &=-\sgn(\hat{U}_e(v)\cdot g_{vv'}\hat{U}_e(v'))\sgn V_4(v)\sgn V_4(v')
    \end{split}
  \end{equation}
  where we recall $\hat{U}_e(v)=\tilde{\ep}g_{vv'}\hat{U}_e(v')$. Because for each face in a 4-simplex there are always two edge bound it. When we flip the sign of five $\hat{U}_e(v)$s in $\sigma_v$, the product $\prod_{e\in f}\mu_e$ is not changed for both internal and boundary faces. $\Box$

  Furthermore as showing in Fig.\ref{fig:boundary}, because of Eq.(\ref{eq:RconHLB}), we can parallel transform three segment vectors $E_l(e_0)$ to $E_l(e_1)$ by using $G_{e_1e_0}\equiv g_{e_1v_1}\cdots g_{v_0e_0}$
  \begin{equation}
    \bigg(\prod_e\mu_e\bigg)E_{pp'}(e_1)=G_{e_1e_0}E_{pp'}(e_0), \quad \forall pp'\in f_l
  \end{equation}
  Therefore the triangle $f_l$ formed by $E_{pp'}(e_0)$ matches in shape with the triangle formed by $E_{pp'}(e_1)$. Since both $E_{pp'}(e_0)$ and $E_{pp'}(e_1)$ are orthogonal to the time gauge vector $u=(1,0,0,0)$, there is a O(3) matrix $g_l$ such that
  \begin{equation}\label{eq:shapM}
    g_l E_{pp'}(e_0)=E_{pp'}(e_1)
  \end{equation}
  These relation gives the restrictions of the boundary data for the spin foam amplitude. We call the boundary condition the non-degenerate \emph{Regge boundary condition}.

  \subsection{Reconstruction theorem}
  In this subsection, we summarize all the discussion in this section as a reconstruction theorem (see also \cite{Conrady:SL2008} for the case of a simplicial manifold without boundary)
  \begin{Theorem}[Reconstruction Theorem]
  Given a set of data $\{j_f,n_{ef},g_{ve}\}$ be a non-degenerate critical configuration which solves Eqs.(\ref{eq:gluingEX}), (\ref{eq:closeEX}), (\ref{eq:PTvv}), and (\ref{eq:SimCstE}) on a simplicial manifold with boundary, there exists a discrete classical Euclidean geometry represented by a set of segment vectors $E_l(v)$ satisfying Eqs.(\ref{eq:inverse}),(\ref{eq:Close}) and (\ref{eq:gluing}) in the bulk, and $E_l(e)$ satisfying Eqs.(\ref{eq:inverseB}),(\ref{eq:CloseB}),(\ref{eq:gluingB}) and (\ref{eq:GaugeB}) on the boundary, such that
  \begin{enumerate}
    \item the semi-geometrical area bivectors $X_f(v)$ and $X_{ef}$ from spin foam stationary points can reconstruct a non-oriented area bivectors of a classical discrete bivectors $A_f(v)$ and $A_{ef}$ up to a global sign $\ep$
        \begin{eqnarray}
           X_f(v)&=\ep A_f(v)&=\ep\frac{1}{2}(E_{l}(v)\wedge E_{l'}(v))\\
           X_{ef}&=\ep A_{ef}&=\ep\frac{1}{2}(E_{l}(e)\wedge E_{l'}(e))
        \end{eqnarray}
        where $l,~l'$ are the segments of triangle $f$. Moveover the segment vectors are totally determined up to an inverse sign $E_{l}\mapsto-E_{l}$. With the segment vectors $E_l(v)$ and $E_l(e)$ we can reconstruct the discrete metric $g_{ll'}$ from them
        \begin{eqnarray}
           g_{ll'}(v)&=&\delta_{IJ}E_l^I(v)E_{l'}^J(v)\\
           g_{ll'}(e)&=&\delta_{IJ}E_l^I(e)E_{l'}^J(e)
        \end{eqnarray}
        The metric is independent of $v$ and $e$ is because the gluing conditions Eqs.(\ref{eq:gluing}) and (\ref{eq:gluingB}). The norm of the bivector $X_f(v)$ is $\gamma j_f$ which is understood as the area of the triangle $f$.
    \item $\mathrm{sgn}V_3(e)$ has to be chosen as a constant $\forall t_e\in\p\Delta$, then the global sign factor $\ep$ is fixed to be $\ep=\mathrm{sgn}V_3(e)$ when we choose the orientations of the triangles such that $\ep_{e_ie_j}(v)\epsilon_{e_me_ne_ke_ie_j}(v)=1$ identically.

    \item $\forall e=(vv')\in\Delta^*, l\in t_e\in\Delta$, the segment vectors in $E_l(v)$ and $E_l(v')$ are related by parallel transformation $g_{vv'}\in$SO(4) associated with edge $e$ up to a sign $\mu_e$
        \begin{equation}
          \mu_e E_{e_1e_2}(v)=g_{vv'}\rhd E_{e_1e_2}(v'), \quad \forall [e_1e_2]\in t_e
        \end{equation}
        $\forall t_e\in\p\Delta$, $t_e\in\sigma_v$, the segment vectors $E_{e_1e_2}(e)$ and $E_{e_1e_2}(v)$ are also related by parallel transformation $g_{ve}\in$SO(4) associated with half edge $(ev)$ up to a sign $\mu_e$
        \begin{equation}
          \mu_e E_{e_1e_2}(v)=g_{ve}\rhd E_{e_1e_2}(e), \quad \forall [e_1e_2]\in t_e
        \end{equation}
        Thus the critical point of $g_{vv'}$ and $g_{ve}$ can be related with SO(4) matrices $\Omega_{vv'}$ and $\Omega_{ve}$ up to the same sign as the one relate $E_{e_1e_2}$s
        \begin{equation}
          g_{vv'}=\mu_e\Omega_{vv'}, \quad g_{ve}=\mu_e\Omega_{ve}
        \end{equation}
        The simplicial complex $\Delta$ can be subdivided into sub-complexes $\Delta_1,\cdots,\Delta_n$ such that (1) each $\Delta_i$ is a simplicial complex with boundary, (2) within each sub-complex $\Delta_i$, $\sgn V_4$ is a constant. Then within each sub-complex $\Delta_i$, the SO(4) matrices $\Omega_{vv'}$ and $\Omega_{ve}$ are the discrete spin connection compatible with the segment vectors $E_{e_1e_2}$s.
    \item Given the boundary triangles $f$ and boundary tetrahedrons $t_e$, in order to have non-degenerate solutions to the equations of motion. The spin foam boundary data $\{j_f, n_{ef}\}$ must satisfy the non-degenerate Regge boundary conditions: (1) For each boundary tetrahedron $t_e$ and its triangles $f$, $\{j_f, n_{ef}\}$ determines 4 triangle normals ${n}_{ef}$ that spans a 3-dimensional subspace. (2) The boundary data are restricted to be shape matched Eq.(\ref{eq:shapM}). (3) The boundary triangulation is consistently oriented such that $\sgn V_3$ is a constant on the boundary. If the Regge boundary condition is satisfied, there are non-degenerate solutions of the equations of motion.
  \end{enumerate}
  \end{Theorem}

  \section{Spin foam amplitude at non-degenerate critical configurations}\label{sec:AsympND}
  The asympotics of the spin foam amplitude is a sum the amplitude evaluated at the critical configurations. In this section, we evaluate the spin foam amplitude at the non-degenerate critical configurations. We show that the spin foam action at a non-degenerate critical configuration is almost a Regge action. As we mentioned in the last section, we subdivide the complex $\Delta$ into sub-complexes $\Delta_1,\cdots,\Delta_n$ such that (1) each $\Delta_i$ is a simplicial complex with boundary, (2) within each sub-complex $\Delta_i$, $\sgn V_4$ is a constant. To study the spin foam (partial-)amplitude $Z_{j_f}(\Delta)$ at a non-degenerate critical configuration $\{j_f, n_{ef}, g_{ve}\}$, we only need to study the amplitude $Z_{j_f}(\Delta_i)$ on the sub-complex $\Delta_i$. The amplitude $Z_{j_f}(\Delta)$ can be expressed as
  \begin{equation}
    Z_{j_f}(\Delta)=\prod_i Z_{j_f}(\Delta_i)
  \end{equation}
  Therefore the following analysis is in one of $\Delta_i$.

  \subsection{Internal faces}
  We first consider an internal faces $f_i$. For an internal face, the action defined in Eq.(\ref{eq:actionE}) can be rewritten in the following way: By Eq.(\ref{eq:gluingEs}), the parallel transportation acting on the coherent state $|n_{ef}\rangle $ gives
  \begin{equation}
    g_{e^{\prime }v}^{\pm }g_{ve}^{\pm }|n_{ef}\rangle =e^{i\phi _{e^{\prime}ve}^{\pm }}|n_{e^{\prime }f}\rangle
  \end{equation}
  Thus the loop holonomy along the boundary of an oriented face $f$ gives
  \begin{equation}
      g_{ev_{k}}^{\pm }g_{v_{k}e_{k}}^{\pm }\cdots g_{e_{1}v}^{\pm }g_{ve}^{\pm}|n_{ef}\rangle=\exp \left( \im\Phi _{f}^{\pm }\right)|n_{ef} \rangle
  \end{equation}
  where $\Phi _{f}^{\pm }\equiv \sum_{v\in f}\phi _{vf}^{\pm }$. This implies the loop holonomy can be written as
  \begin{equation}
    G_{f}^{\pm }\left( e\right) \equiv g_{ev_{k}}^{\pm }g_{v_{k}e_{k}}^{\pm}\cdots g_{e_{1}v}^{\pm }g_{ve}^{\pm }=\exp \left(\im\Phi _{f}^{\pm}\hat{X}^{\pm}_{ef}\right)
  \end{equation}
  where $\hat{X}^{\pm}_{ef}=\im (n_{ef})_i\sigma^i$ as defined in Eq.(\ref{eq:PTsa}).
  Then the action defined in Eq.(\ref{eq:actionE}) turns into
  \begin{equation}\label{eq:SfPhipm}
    \begin{split}
      S_{f} &=\sum_{v\in f}\sum_{\pm }2j_{f}^{\pm }\ln \left\langle n_{ef}\left\vert g_{ev}^{\pm }g_{ve^{\prime }}^{\pm }\right\vert n_{e^{\prime }f}\right\rangle\\
      &=\sum_{\pm }2ij_{f}^{\pm }\Phi _{f}^{\pm }
    \end{split}
  \end{equation}

By using the following identity
  \begin{equation}\label{eq:trick}
    \exp \left( \im\Phi _{f}^{\pm }\right) =\tr\left( \frac{1}{2}\exp \left( \im\Phi_{f}^{\pm }n_{ef}^{i}\sigma _{i}\right) \left( 1+n_{ef}^{i}\sigma_{i}\right) \right)
  \end{equation}
  we can rewrite the action in the following form
  \begin{equation}
    S_{f}=\sum_{\pm }2j_{f}^{\pm }\ln \left[ \tr\left( \frac{1}{2}G_{f}^{\pm }(e) \left( 1+\hat{X}^{\pm}_{ef}\right) \right) \right]
  \end{equation}
  Then let us use the parallel transformation of $G_{f}^{\pm }(e)$ and $\hat{X}^{\pm}_{ef}$ to take them to the nearest vertex $v$.
  \begin{eqnarray}
    G_{f}^{\pm }(v) &\equiv& g_{ve}^{\pm }G_{f}^{\pm }(e)g_{ev}^{\pm }\\
    \hat{X}^{\pm}_{f}(v)&=&g^{\pm}_{ve}\hat{X}^{\pm}_{ef}g^{\pm}_{ev}
  \end{eqnarray}
  Because the trace does not change under the parallel transformation, then the action becomes
  \begin{equation}\label{eq:Ssf}
    S_{f}=\sum_{\pm }2j_{f}^{\pm }\ln \left[ \tr\left( \frac{1}{2}G_{f}^{\pm }(v) \left( 1+ \hat{X}^{\pm}_{f}(v) \right) \right) \right]
  \end{equation}

  We would like to find the relation between Eq.(\ref{eq:Ssf}) and Regge action. Recalling Eq.(\ref{eq:RconstAfX}) and Eq.(\ref{eq:RconHL}) we have the parallel transportation of $X_f(v)$ and $E_{pp'}(v)$ under the loop holonomy $G_f(v)$
  \begin{equation}\label{eq:PTLX}
    G_f(v)\rhd X_f(v)=X_f(v)
  \end{equation}
  \begin{equation}\label{eq:PTLE}
    \begin{split}
      G_f(v)\rhd E_{pp'}(v)&\equiv\exp\bigg(\im\pi\sum_{e\in f}n_e\bigg)E_{pp'}(v)\\
      &=\cos\bigg(\pi\sum_{e\in f}n_e\bigg)E_{pp'}(v)
    \end{split}
  \end{equation}
  where $\exp(\im\pi\sum_{e\in f}n_e)\equiv\prod_{e\in f}\mu_e$, $pp'\in f$. These equations imply that the loop holonomy $G_f(v)$ gives a rotation in the plane orthogonal to the 2-plane determined by $X_f(v)$, i.e. in the plane of $\star X_f(v)$. Then we can explicit write the loop holonomy as
  \begin{equation}\label{eq:LHsf}
    G_f(v)=\exp\left( \star\hat{A}_f(v)\theta_f\right)\exp\bigg(\pi\sum_{e\in f}n_e\hat{A}_f(v)\bigg)
  \end{equation}
  where $\hat{A}_f(v)=A_f(v)/|A_f(v)|$. The transformation Eq.(\ref{eq:PTLE}) can be shown in the following way. Give a bivector $A_f(v)=(E_1(v)\wedge E_2(v))/2$, we choose two orthogonal basis $e_1=(0,0,1,0)$, $e_2=(0,0,0,1)$ in the plane of $A_f(v)$, then $\hat{A_f}(v)=e_1\wedge e_2$. We can show that on the 2-plane of $A_f(v)$
  \begin{equation}
    \exp\bigg(\pi\sum_{e\in f}n_e\hat{A_f}(v)\bigg)=\cos\bigg(\pi\sum_{e\in f}n_e\bigg)\mathds{1}
  \end{equation}
is a $\pi$-rotation.

Because the relation between the spin foam variable $g_{vv'}$ and spin connection $\Omega_{vv'}$ Eq.(\ref{eq:RconHL}), we can get the loop spin connection $\Omega_f(v)$
  \begin{equation}\label{eq:omegaf}
    \Omega_f(v)=\exp\bigg(\im\pi\sum_{e\in f}n_e\bigg)\exp\bigg( \star\hat{A}_f(v)\theta_f+\pi\sum_{e\in f}n_e\hat{A}_f(v)\bigg)
  \end{equation}
From the discussion in Section \ref{sec:DcG}, we have the geometrical interpretation of the parameter $\theta_f$
\begin{equation}
\theta_f=\sgn V_4 \Theta_f
\end{equation}
where $\Theta_f$ is the deficit angle in Regge calculus.

  Based on the relation between SO(4) and its self-dual and anti-self-dual decomposition Eq.(\ref{eq:so42su2}) and Eq.(\ref{eq:SO4toSU2}), the self-dual and anti-self-dual loop holonomy $G^{\pm}_f(v)$ are
  \begin{equation}
    \begin{split}
      G^{\pm}_f(v)&=\exp\left[\frac{\im}{2}\ep\left(\pi\sum_{e\in f}n_e\mp\sgn V_4\Theta_f\right)\hat{X}_f^{\pm}\right]\\
      &=\exp(\im\Phi_f^{\pm}\hat{X}_f^{\pm})
    \end{split}
  \end{equation}

  We take $\Phi^{\pm}_f$ defined above into Eq.(\ref{eq:SfPhipm}), then we can get the asymptotic action for interior faces
  \begin{equation}
    S_{f_i}=-\im\ep\sum_{f_i}\gamma j_{f_i} \sgn V_4\Theta_{f_i}+\im \ep \pi\sum_{f_i}j_{f_i} \left(\sum_{e\in f}n_e\right)
  \end{equation}
  where $\gamma j_{f_i}$ is the area of triangle $f_i$ and the first term $\sgn V_4\sum_{f_i}\gamma j_{f_i} \Theta_{f_i}$ is the Regge action for discrete GR when $\sgn V_4$ is a constant.

  \subsection{Boundary faces}
  Now we consider the action for boundary face $f$. Giving a boundary face $f$ as shown in Fig.\ref{fig:boundary}, together with the gluing condition Eq.(\ref{eq:gluingEs}), we obtain
  \begin{equation}
    g_{e_1v_{1}}^{\pm }\cdots g_{v_0e_0}^{\pm}|n_{e_0f}\rangle=\exp \left( \im\Phi _{e_1e_0}^{\pm }\right)|n_{e_1f} \rangle
  \end{equation}
  This implies that the holonomy $G_{e_1e_0}^{\pm}\equiv g_{e_1v_{1}}^{\pm }\cdots g_{v_0e_0}^{\pm}$ can be written as
  \begin{equation}
    G_{e_1e_0}^{\pm}=g(n_{e_1f})\exp\left[\im\Phi_{e_1e_0}^{\pm}\sigma_z\right]g^{-1}(n_{e_0f})
  \end{equation}
  where the SU(2) group element $g(n)$ is given by
  \begin{equation}
    g(n)=\ket{n}\bra{z}+\ket{Jn}\bra{Jz}
  \end{equation}
  in spin-$\frac{1}{2}$ representation. In spin-1 representation, it rotates $z=(0,0,1)$ to the 3-vector $n$. We can also consider $g(n)$ as a rotation from the reference frame at $f$ to the reference frame in the tetrahedron $t_e$.

The action of $f_l$ can be written as
  \begin{equation}
    S_{f} =\sum_{\pm }2ij_{f}^{\pm }\Phi _{e_1e_0}^{\pm }
  \end{equation}
which can be rewritten using Eq.(\ref{eq:trick}) as
  \begin{equation}\label{eq:SsfB}
    S_{f}=\sum_{\pm }2j_{f}^{\pm }\ln \left[ \tr\left( \frac{1}{2}g^{-1}(n_{e_1f})G_{e_1e_0}^{\pm}g(n_{e_0f}) \left( 1+\hat{X}^{\pm}_z\right) \right) \right]
  \end{equation}
  where $\hat{X}^{\pm}_z= \sigma_z$.

To find the relation between Eq.(\ref{eq:SsfB}) and the Regge action. We redefine the segment vectors $E_l(e_i)$ as
  \begin{equation}
    \tilde{E}_{pp'}(e_i)\equiv g^{-1}(n_{e_if})\rhd E_{pp'}(e_i),\quad \forall pp'\in f
  \end{equation}
  Since $E_{pp'}(e_i)$ is orthogonal to $n_{e_i}$, we can get $\tilde{E}_{pp'}(e_i)$ is orthogonal to $z$. For $p,p'$ vertices of the triangle $f$, $\tilde{E}_{pp'}(e_0)$ and $\tilde{E}_{pp'}(e_1)$ must be related by a rotation in the plane of $f$. Here we gauge fix this rotation to be identity, i.e. $\tilde{E}_{pp'}(e_1)= \tilde{E}_{pp'}(e_0)\equiv E_{pp'}(f)$. Then recall the parallel transportation of the bivector $X_{e_1f}=G_{e_1e_0}\rhd X_{e_0f}$, and Eq.(\ref{eq:RconHL}), Eq.(\ref{eq:RconHLB}) we have the following relations ($p,p'$ are vertices of the triangle $f$)
  \begin{eqnarray}
      \tilde{G}_{e_1e_0} \rhd {E}_{pp'}(f)&=&\left(\prod_{e\in f_l}\mu_e\right) {E}_{pp'}(f) \nonumber\\
      &=&\cos\bigg(\pi\sum_{e\in f}n_e\bigg){E}_{pp'}(f)
  \end{eqnarray}
  where $\tilde{G}_{e_1e_0}\equiv g^{-1}(n_{e_1f})G_{e_1e_0}g(n_{e_0f})$. The above equation implies that the parallel transportation $g^{-1}(n_{e_1f})G_{e_1e_0}g(n_{e_0f})$ has the following form (see Section\ref{sec:DcG})
  \begin{equation}\label{eq:LHsfB}
    \tilde{G}_{e_1e_0}=\exp\bigg(\star\hat{{A}}_f(f)\theta_f^B+\pi\sum_{e\in f}n_e{\hat{A}}_f(f)\bigg)
  \end{equation}
where $\theta^B_f$ is the parameter of the dihedral rotation, and $\hat{A}_f(f)=\ep \hat{X}_z$. From Eq.(\ref{eq:RconHL}), Eq.(\ref{eq:RconHLB}) and Eq.(\ref{eq:LHsfB}), we have
  \begin{equation}\label{eq:omegaee}
    \Omega_{e_1e_0}=\exp\bigg(\im\pi\sum_{e\in f}n_e\bigg)g(n_{e_1f})\tilde{G}_{e_1e_0}g^{-1}(n_{e_0f})
  \end{equation}

From the discussion in Section\ref{sec:DcG} and \cite{Han:2011AsLorentz}, The parameter $\theta^B_f$ relates to the dihedral angle $\Theta_f^B$ between the two tetrahedrons $t_{e_0}$ and $t_{e_1}$ by
\begin{equation}
\theta^B_f=\Theta^B_f\sgn V_4(v).
\end{equation}

  Then taking Eq.(\ref{eq:LHsfB}) back to Eq.(\ref{eq:SsfB}), we can get for any boundary faces
  \begin{equation}
    S_{f_e}=-\im\ep\sum_{f_e}\gamma j_{f_e} \sgn V_4\Theta_{f_e}^B+\im \ep\pi\sum_{f_e}j_{f_e} \left(\sum_{e\in f}n_e\right)
  \end{equation}

  \subsection{Asymptotic non-degenerate amplitude}
  As we have shown above, for any set of non-degenerate solutions $\{j_f,n_{ef}, g_{ve}\}$ of Eq.(\ref{eq:gluingEX}) and Eq.(\ref{eq:closeEX}) together with Eq.(\ref{eq:orientEs}), Eq.(\ref{eq:SimCstE}), Eq.(\ref{eq:PTev}), we can always construct a non-degenerate discrete geometry with a global sign ambiguity $\ep$.

  We briefly summarize the results we get so far. For a given non-degenerate critical configuration $\{j_f,n_{ef}, g_{ve}\}$, we can reconstruct the discrete geometric variables $E_l$ and $U^e$.
  \begin{itemize}
    \item $\forall v\in \Delta^*$, we can reconstruct a bivector geometry of 4-simplex. Given any semi-geometrical bivector $X_f(v)$ from the critical configuration, there is a non-oriented bivector $A_f(v)=(E_l(v)\wedge E_{l'}(v))/2$ in discrete geometry such that
        \begin{equation}
          X_f(v)=\ep A_f(v)
        \end{equation}
        where $\ep$ is a global sign on the entire simplicial complex $\Delta$.
    \item $\forall e\in\Delta^*$, we can associate a spin connection $\Omega_e$ (when $\sgn V_4(v)=\sgn V_4(v')$) by the on-shell $g_{vv'}$ up to a sign $\mu_e$
        \begin{equation}
           g_{vv'}=\mu_e\Omega_e
        \end{equation}
        where $v$ and $v'$ are the end points of $e$.
    \item $\forall e\in \p\Delta^*$, we can construct the segment vectors $E_l(e)$ such that giving any semi-geometrical bivector $X_f(e)$ from the critical configuration, we can find a non-oriented bivector $A_f(e)=(E_l(e)\wedge E_{l'}(e))/2$ in discrete geometry on the boundary that
        \begin{equation}
          X_f(e)=\ep A_f(e)
        \end{equation}
  \end{itemize}
A non-degenerate critical configuration $(j_f,g_{ve},n_{ef})$ specifies uniquely a set of variables $(g_{l_1l_2},n_e,\ep)$, which include a discrete metric and two types of sign factors.

Given a critical configuration $(j_f,g_{ve},n_{ef})$ in general, we can divide the triangulation $\Delta$ into sub-triangulations $\Delta_1,\cdots,\Delta_n$, where each of the sub-triangulations is a triangulation with boundary, with a constant $\mathrm{sgn}(V_4(v))$. On each of the sub-triangulation $\Delta_i$, we add the on-shell actions of internal and boundary faces together, we have
  \begin{equation}\label{eq:actionNondeg}
    \begin{split}
      S_f(g_{l_1l_2},n_e,\ep)&|_{\mathrm{Non-deg}}=\sum_{f_i}S_{f_i}+\sum_{f_e}S_{f_e}\\
      &=-\im\ep\sgn V_4\bigg(\sum_{f_i}\gamma j_{f_i} \Theta_{f_i}+\sum_{f_e}\gamma j_{f_e} \Theta_{f_e}^B\bigg)\\
      &\quad +\im \ep\pi \bigg(\sum_{e}n_e\sum_{f\in e}j_{f}\bigg)
    \end{split}
  \end{equation}
Here $\Theta_{f_i}$ and $\Theta^B_{f_e}$ are the deficit angle and the dihedral angle respectively, which are determined only by the discrete metric $g_{ll'}$. Moreover $\forall e\in \Delta^*$ $\sum_{f\in e}j_{f}$ is an integer. It contribute an overall sign when we exponentiate $S_f(g_{l_1l_2},n_e,\ep)$.

We say a spin configuration $j_f$ is Regge-like if there exist the critical configurations solving the equation of motion, which is non-degenerate everywhere. Given a collection of Regge-like spins $j_f$ for each 4-simplex, the discrete metric $g_{\ell_1\ell_2}(v)$ is uniquely determined for the simplex. Furthermore since the areas $\gamma j_f$ are Regge-like, There exists a discrete metric $g_{l_1l_2}$ in the entire bulk of the triangulation, such that the neighboring 4-simplicies are consistently glued together, as we constructed previously. This discrete metric $g_{l_1l_2}$ is obviously unique by the uniqueness of $g_{l_1l_2}(v)$ at each vertex. Therefore given the partial-amplitude $Z_{j_f}(\Delta)$ with a specified Regge-like $j_f$, all the critical configurations $(j_f,g_{ve},n_{ef})$ with the same Regge-like $j_f$ correspond to the same discrete metric $g_{\ell_1\ell_2}$, provided a Regge boundary data. The critical configurations from the same Regge-like $j_f$ is classified in the next section.

  As a result, for any Regge-like configurations $j_f$ and a Regge boundary data $n_{ef_e}$, the amplitude $Z_{j_f}(\Delta)|_{\mathrm{Non-deg}}$ has the following asymptotic behavior
  \begin{equation}\label{eq:nondegA}
    \begin{split}
      Z_{j_f}(&\Delta)|_{\mathrm{Non-deg}}\sim \sum_{x_c} C(x_c)\left[1+\mathcal{O}\left(\frac{1}{\lambda}\right)\right]\\
      &\times  \exp\lambda\sum_{\Delta_i}\bigg[-\im\ep\sgn V_4\bigg(\sum_{f_i}\gamma j_{f_i} \Theta_{f_i}+\sum_{f_e}\gamma j_{f_e} \Theta_{f_e}^B\bigg)\\
      &\quad\quad\quad\quad+\im \ep\pi \bigg(\sum_{e}n_e\sum_{f\in e}j_{f}\bigg)\bigg]
    \end{split}
  \end{equation}
  where $x_c$ stands for the non-degenerate critical configurations $(j_f,g_{ve},n_{ef})$ and $C(x_c)$ is given by the follows
  \begin{equation}\label{eq:Cx0} C(x_c)=a(x_c)\left(\frac{2\pi}{\lambda}\right)^{\frac{r(x_c)}{2}}\frac{\ex^{\im\mathrm{Ind}H'(x_c)}}{\sqrt{|\det_rH'(x_c)|}}
  \end{equation}
  where $H(x_c)$ is the Hessian matrix of the action $S_f$ and $H'(x_c)$ is the invertible restriction on $\mathrm{ker}H(x_c)^{\bot}$; $r(x_c)$ is the rank of Hessian matrix. The on-shell action on exponential gives the Regge action up to the sign factor $\sgn V_4\big|_{\Delta_i}$ of the oriented 4-volume. However if we recall the difference between the Einstein-Hilbert action and Palatini action
\begin{eqnarray}
\mathcal{L}_{EH}&=&R\ \underline{\ep}=\mathrm{sgn}\det(e_\mu^I)*\![e\wedge e]_{IJ}\wedge R^{IJ}\nonumber\\
&=&\mathrm{sgn}\det(e_\mu^I)\mathcal{L}_{Pl}
\end{eqnarray}
where $\mathcal{L}_{EH}$ and $\mathcal{L}_{Pl}$ denote the Lagrangian densities of Einstein-Hilbert action and Palatini action respectively, and $\underline{\ep}$ is a chosen volume form compatible with the metric $g_{\mu\nu}=\eta_{IJ}e_\mu^I e_\nu^J$. Since the Regge action is a discretization of the Einstein-Hilbert action, we may consider the resulting action from the asymptotics is a discretization of the Palatini action with the connection compatible with the tetrad.

   \section{Parity Inversion}\label{sec:parity}
  Given a tetrahedron $t_e$ associated with spins $j_{f_1},\cdots, j_{f_4}$, we know that the set of four normals $n_{ef_{1}},\cdots,n_{ef_{4}}\in S^2$, modulo diagonal SO(3) rotation, is equivalent to the shape of $t_e$, if closure condition is satisfied \cite{Conrady:2009px}\cite{Bianchi:2010gc}. Given a set of non-degenerate solutions and configurations $\{j_f, g_{ve},n_{ef}\}$, as discussed above, the Regge-like spin configuration $j_f$ determines a discrete metric $g_{ll'}$, which determines the shape of all the tetrahedrons in $\Delta$. The diagonal SO(3) rotation of $n_{ef_{1}},\cdots,n_{ef_{4}}$ is also a gauge transformation of the spin foam action. Thus the gauge equivalence class of the critical configurations $\{j_f, n_{ef}, g_{ve}\}$ with the same Regge-like spins $j_f$ must have the same set of $n_{ef}$. The degrees of freedom of the non-degenerate critical configurations are the freedom of the variables $g_{ve}$ when we fix a Regge-like $j_f$. The degrees of freedom of $g_{ve}$ are encoded in the 4-simplex geometry. Given a set of data $\{j_f, n_{ef}\}$, the non-degenerate critical configurations within each 4-simplex are completely classified \cite{Barrett:2009gg}\cite{Barrett:2009mw} and are related by parity transformation.

  Given a set of non-degenerate solutions and configurations $\{j_f, g_{ve},n_{ef}\}$, we can generate many other sets of solutions and configurations $\{j_f, \tilde{g}_{ve},n_{ef}\}$. As discussed in \cite{Barrett:2009gg}, the two solutions $g_{ve}$ and $\tilde{g}_{ve}$ are related by local parity in some 4-simplices. In Euclidean theory, within a 4-simplex $\sigma_v$, if $g_{ve}=(g_{ve}^+,g_{ve}^-)$ is a solution of equations of motion, $\tilde{g}_{ve}=(g_{ve}^-,g_{ve}^+)$ is also a solution of the same equations. The semi-geometric variables generated by $g_{ve}$ and $\tilde{g}_{ve}$ are related by local parity transformation, since
  \begin{eqnarray}
    N^e(v)_I\sigma^I_E&\equiv& g_{ve}^- u_I\sigma^I_E (g_{ve}^+)^{-1}=g_{ve}^-(g_{ve}^+)^{-1}\\
    \tilde{N}^e(v)_I\sigma^I_E&\equiv& g_{ve}^+ u_I\sigma^I_E (g_{ve}^-)^{-1}=g_{ve}^+(g_{ve}^-)^{-1}
  \end{eqnarray}
  Then
  \begin{equation}
    \begin{split}
      \tilde{N}^e(v)_I\sigma^I_E&=(N^e(v)_I\sigma^I_E)^{\dag}=N^e_0(v)\mathds{1}-\im N^e_i(v)\sigma^i\\
      &=(\mathbf{P}N^e(v))_I\sigma^I_E
    \end{split}
  \end{equation}
  where $\mathbf{P}$ is the parity operator on Euclidean vector space.

  Then let us look at the relation between semi-geometric bivectors $X_f(v)$ and $\tilde{X}_f(v)$, where $\tilde{X}_f(v)$ is the bivector defined by using $\tilde{g}_{ve}$. We have the relations $\tilde{X}^\pm_f(v)=X^\mp_f(v)$
  \begin{equation}
  \begin{split}
    X_i^{\pm}(v)&=\frac{1}{2}\epn_i^{~jk}X_{jk}(v)\pm X_{i0}(v)\\
    X_i^{\pm}(v)&=\frac{1}{2}\epn_i^{~jk}\tilde{X}_{jk}(v)\mp \tilde{X}_{i0}(v)
  \end{split}
  \end{equation}
  Then we can easily get $X_{i0}(v)=-\tilde{X}_{i0}(v)$ and $X_{jk}(v)=\tilde{X}_{jk}(v)$, i.e.
  \begin{equation}\label{eq:paX}
    \tilde{X}_f(v)=\mathbf{P}\rhd X_f(v)
  \end{equation}

  Reminding Eq.(\ref{eq:Xf}), we can get
  \begin{equation}
  \begin{split}
    \tilde{X}_f(v)&=\tilde{\alpha}_{ee'}(v)\star(\tilde{N}^e(v)\wedge\tilde{N}^{e'}(v))\\
    &=-\tilde{\alpha}_{ee'}(v)\mathbf{P}\rhd \star(N^e(v)\wedge N^{e'}(v))\\
    &=-\frac{\tilde{\alpha}_{ee'}(v)}{\alpha_{ee'}(v)}\mathbf{P}\rhd X_f(v)
  \end{split}
  \end{equation}
  Then recall Eq.(\ref{eq:paX}) we get
  \begin{equation}
    \tilde{\alpha}_{ee'}(v)=-\alpha_{ee'}(v)
  \end{equation}
  Because $\beta_{ee'}(v)=\alpha_{ee'}(v)\ep_{ee'}(v)$,
  \begin{equation}
    \tilde{\beta}_{ee'}(v)=-\beta_{ee'}(v)
  \end{equation}
  Then $\beta_{ii}(v)=\tilde{\beta}_{ii}(v)$ since $\sum_j\beta_{ij}N^{e_j}=0$. Based on this and the definition of $\beta _{i}(v)\equiv\beta _{ij_0}(v)/\sqrt{|\beta _{j_0j_0}(v)|}$ we can get
  \begin{equation}\label{eq:Pbeta}
    \tilde{\beta}_{i}(v)=-\beta_{i}(v)
  \end{equation}
  As in subsection \ref{subsec:DGdeSS}, use Eq.(\ref{eq:defV4}) we can get
  \begin{equation}\label{eq:PV}
    \tilde{V}_4(v)=-V_4(v)
  \end{equation}
  The minus sign is from the fact that $\det\mathbf{P}=-1$. Then the global signs $\ep(v)=\mathrm{sgn}(\beta_{j_0j_0})\mathrm{sgn}(V_4)$ are the same
  \begin{equation}
    \tilde{\ep}(v)=\ep(v)
  \end{equation}
  This result shows that when we change $\sigma_v$ into its parity one $\tilde{\sigma}_v$, the global sign stay invariant.

  The fact that the local parity change the sign of the 4-volume of the 4-simplex leads to some interesting consequences. First of all, given any critical configuration $\{j_f,g_{ve}, n_{ef}\}$ with a Regge-like spin configuration $\{j_f\}$, we can always subdivide the triangulation $\Delta$ into sub-complexes. Each of the sub-complex has a constant $\sgn V_4$. Now we understand that the local parity transforms a configuration $\{j_f,g_{ve}, n_{ef}\}$ to a new configuration $\{j_f,\tilde{g}_{ve}, n_{ef}\}$, which may have different subdivision according to $\sgn V_4$. On the other hand, for each subdivision with a critical configuration $x_c=\{j_f, g_{ve}, n_{ef}\}$, there is always another critical configuration $\tilde{x}_c=\{j_f, \tilde{g}_{ve}, n_{ef}\}$ obtained from the former one by a global parity, which leaves the subdivision unchange but changes the 4-volume sign within each sub-complex. Thus the global parity changes the spin foam action at the non-degenerate stationary configuration into its opposite, i.e.
  \begin{equation}
    S(x_c)=-S(\tilde{x}_c).
  \end{equation}
Note that the deficit angle, dihedral angle, and $\sum_{e\subset\partial f}n_e$ are unchanged under the global parity, which is shown in the follows.

  Then let us get the relation between segment vector $\tilde{E}_l(v)$ and $E_l(v)$. Because Eqs.(\ref{eq:NtoU}) and (\ref{eq:Pbeta}), we can get
  \begin{equation}
    \tilde{U}^e(v)=-\mathbf{P}U^e(v)
  \end{equation}
  Then based on Eqs.(\ref{eq:UtoE}) and (\ref{eq:PV}), we can have
  \begin{equation}
    \tilde{E}_l(v)=-\mathbf{P}E_l(v)
  \end{equation}
  From the above discussion we can find that the local and global parity inversion $\tilde{E}_l(v)=-\mathbf{P}E_l(v)$ does not change the discrete metric $g_{ll'}=\delta_{IJ}E_l^I(v)E_{l'}^J(v)$. Thus the parity configuration $\tilde{x}_c$ gives the same discrete geometry as $x_c$, and only make an O(4) gauge transformation to the segment vectors. The matrix $\Omega_{vv'}$ is uniquely determined by $E_l(v)$ and is a spin connection as long as $\sgn V_4(v)=\sgn V_4(v')$, as shown in Section \ref{sec:AsympND}. The global parity transformation does not change the subdivisions but flip the signs of $\sgn V_4$ in each sub-complex. Given a spin connection $\Omega_{vv'}$ in a subdivision, the parity one $\tilde{\Omega}_{vv'}$ is
  \begin{equation}
    \tilde{\Omega}_{vv'}=\mathbf{P}\Omega_{vv'}\mathbf{P}
  \end{equation}
  since $\tilde{\Omega}_{vv'}\tilde{E}_{ee'}(v')=\tilde{E}_{ee'}(v)$. One can check that for a 4-vector $V^I$, $\tilde{g}\mathbf{P}V=\mathbf{P}g V$. Then from
  \begin{equation}
    \begin{split}
      \tilde{g}_{vv'}\tilde{E}_{ee'}(v')&=-\tilde{g}_{vv'}\mathbf{P}E_{ee'}(v')=-\mathbf{P}{g}_{vv'}E_{ee'}(v')\\
      &=-\mu\mathbf{P}E_{ee'}(v)=\mu\tilde{E}_{ee'}(v)
    \end{split}
  \end{equation}
  and $\tilde{g}_{vv'}=\tilde{\mu}_e\tilde{\Omega}_{vv'}$, we find that $\mu_e$ is invariant under the global parity transformation
  \begin{equation}
    \tilde{\mu}_e=\mu_e
  \end{equation}

  Now let us consider a boundary edge. In the case $t_e$ is a boundary tetrahedron, the parity transform the segment vectors $E_l(v)$ as $\tilde{E}_l(v)=-\mathbf{P}E_l(v)$ at vertex $v$, while leaving the boundary segment vectors $E_l(e)$ invariant. Therefore the spin connection $\tilde{\Omega}_{ve}\in$SO(4) is uniquely determined by
  \begin{equation}
    \tilde{\Omega}_{ve}\tilde{E}_{pp'}(e)=\tilde{E}_{pp'}(v), \quad \forall pp'\in t_e
  \end{equation}
  Then the relation between the spin connection $\Omega_{ve}$ before parity transformation and $\tilde{\Omega}_{ve}$ is
  \begin{equation}
    \tilde{\Omega}_{ve}=-\mathbf{P}\Omega_{ve}\mathbf{T}
  \end{equation}
  where $\mathbf{T}=\mathrm{diag}(-1,1,1,1)$ is the time reversal keeps the spatial vectors $E_l(e)$ unchanged. Then because of $\tilde{g}E(e)=-\mathbf{P}gE(e)$ for spatial vector $E(e)$, we have
  \begin{equation}
    \tilde{g}_{ve}E_{pp'}(e)=-\mu_e\mathbf{P}E_{pp'}(v)=\mu_e \tilde{E}_{pp'}
  \end{equation}
  Then the same as before, we have
  \begin{equation}
    \tilde{\mu}_e=\mu_e
  \end{equation}
  Thus for both interior and exterior faces, the product $\prod_{e\in f}\mu_e$ is invariant under the global parity transformation, i.e.
  \begin{equation}
    \prod_{e\in f}\mu_e=\prod_{e\in f}\tilde{\mu}_e
  \end{equation}
  If we write $\mu_e=\exp(\im\pi n_e)$ and $\tilde{\mu}_e=\exp(\im\pi \tilde{n}_e)$ as before, then we can set
  \begin{equation}
    \sum_{e\in f}n_e=\sum_{e\in f}\tilde{n}_e.
  \end{equation}

  Then let us consider the loop spin connection $\tilde{\Omega}_f$ of an internal face $f$. Based on the discussion about the relation between $\Omega_{vv'}$ and $\tilde{\Omega}_{vv'}$, we have
  \begin{equation}
    \tilde{\Omega}_f(v)=\mathbf{P}\Omega_f(v)\mathbf{P}
  \end{equation}
  Recall Eq.(\ref{eq:omegaf}), we write down the spin connection $\tilde{\Omega}_f(v)$ as
  \begin{equation}
    \begin{split}
      \tilde{\Omega}_f(v)&=\exp\bigg(\im\pi\sum_{e\in f}\tilde{n}_e\bigg)\\
      &\quad\times\exp\bigg( \sgn(\tilde{V}_4)\star\hat{\tilde{A}}_f(v)\tilde{\Theta}_f+\pi\sum_{e\in f}\tilde{n}_e\hat{\tilde{A}}_f(v)\bigg)
    \end{split}
  \end{equation}
  From the relations $\sgn V_4=-\sgn \tilde{V}_4$, $\sum_{e\in f}n_e=\sum_{e\in f}\tilde{n}_e$, $\mathbf{P}\star A_f=-\star \tilde{A}_f$ and $\mathbf{P} A_f=\tilde{A}_f$ we get
  \begin{equation}
    \Theta_f=\tilde{\Theta}_f
  \end{equation}
  which is consistent with the fact that the deficit angle $\Theta_f$ is determined only by the metric $g_{ll'}$ which is invariant under the parity transformation.

  For the holonomy $\tilde{\Omega}_{e_0e_1}$ of a boundary face $f$, the relation between $\tilde{\Omega}_{e_0e_1}$ and $\Omega_{e_0e_1}$ is
  \begin{equation}
    \tilde{\Omega}_{e_0e_1}=\mathbf{T}\Omega_{e_0e_1}\mathbf{T}
  \end{equation}
  As before reminding Eq.(\ref{eq:omegaee}), we can get
  \begin{equation}
    \Theta_f^B=\tilde{\Theta}_f^B
  \end{equation}
  which is consistent with the fact that the dihedral angle $\Theta_f^B$ is determined by the metric $g_{ll'}$ which is invariant under the parity transformation.

Among all the critical configurations $\{j_f, g_{ve}, n_{ef}\}$ with the same Regge-like $j_f$, there exists only two critical configurations such that the signs of the oriented 4-volumes are the same for all the 4-simplex $\sigma_v$ in $\Delta$, while the two configurations are related by a global parity. For any critical configuration, it leads to the subdivision of the triangulation, where each sub-complex has a constant volume sign of the 4-simplexes. As we discussed above, we can always make a certain local/global parity transformation to flip the volume sign within some certain sub-complexes, which constructs the configurations such that the volume sign is constant on the entire simplicial complex.

  \section{Degenerate critical configurations}\label{sec:AsympD}
  In this section, we discuss the degenerate critical configurations. The degeneracy means that there exists  4 different edges $ e_i, e_j, e_k, e_l$ connecting vertex $v$, with a degenerate critical configuration $\{j_f, g_{ve}, n_{ef}\}$, $N^{e}(v)= g_{ve}\rhd u$ satisfy
  \begin{equation}\label{eq:degN}
    \det(N^{e_i}(v),N^{e_j}(v),N^{e_k}(v),N^{e_l}(v))=0.
  \end{equation}

  \subsection{Classification}
  The Lemma 3 in \cite{Barrett:2009gg} shows that within each 4-simplex, all five normals $N_e(v)$ from a degenerate critical configuration $\{j_f, g_{ve}, n_{ef}\}$ are parallel and more precisely $N_e(v) = u = (1,0,0,0)$ once we fix the gauge. This result implies that the self-dual and anti-self-dual parts of SO(4) group element $g_{ve}$ are the same, since
  \begin{equation}
    (g_{ve}u)_I\sigma^I_E=g_{ve}^-u_I\sigma^I_E(g_{ve}^+)^{-1}=u_I\sigma^I_E
  \end{equation}
  In the following discussion of this subsection, if $g_{ve}^-$ and $g_{ve}^+$ are the same, we denote them as $h_{ve}\equiv g_{ve}^-=g_{ve}^+$.

  There are two types of degenerate solutions of the equations of motion. We call them Type-A and Type-B respectively.
  \begin{itemize}
    \item[Type-A] Here the Type-A degenerate configurations are from the non-degenerate configurations. For a Regge-like $\{j_f\}$, as discussed before, we can always solve the equations of motion to get a non-degenerate critical configuration $\{j_f, n_{ef}, g_{ve}\}$ such that we can reconstruct a non-degenerate classical discrete geometry. While as discussed in \cite{Barrett:2009gg}, the equations of motion Eqs.(\ref{eq:gluingEs}),(\ref{eq:closeEs}) in fact coincide with the equations from SU(2) BF theory. For Regge-like $\{j_f\}$, within a 4-simplex, the equations of motion has two different groups of SU(2) solutions $g_{ve}^+$ and $g_{ve}^-$. In addition to the non-degenerate solutions $(g_{ve}^+,g_{ve}^-)$ and $(g_{ve}^-,g_{ve}^+)$, they imply two degenerate SO(4) solutions $(g_{ve}^+,g_{ve}^+)$ and $(g_{ve}^-,g_{ve}^-)$. We call the degenerate configurations $\{j_f, n_{ef}, g_{ve}\}$ defined in this way within all simplices a Type-A configurations.
    \item[Type-B] The equations of motion Eqs.(\ref{eq:gluingEs}),(\ref{eq:closeEs}) may only have one group of SU(2) solutions within a 4-simplex, i.e. we can only find a single SO(4) solution $g_{ve}=(h_{ve},h_{ve})$ \cite{Barrett:2009gg}. We call the configurations and solutions $\{j_f, n_{ef}, g_{ve}\}$ defined in this way within all simplices a Type-B configurations.

  \end{itemize}

  \subsection{Type-A asymptotics}
 The Type-A degenerate configurations are constructed from the non-degenerate critical configurations, which has $g_{ve}^+\neq g_{ve}^-$ in all 4-simplices, as in Section\ref{sec:AsympND}.

 Given two Type-A degenerate solutions $(g_{ve}^+,g_{ve}^+)$ and $(g_{ve}^-,g_{ve}^-)$, we canonically associate the solution $(g_{ve}^+,g_{ve}^+)$ to the non-degenerate solution $(g_{ve}^+,g_{ve}^-)$ for the geometric interpretation, while we associate canonically the other solution $(g_{ve}^-,g_{ve}^-)$ to $(g_{ve}^-,g_{ve}^+)$. Therefore $(g_{ve}^+,g_{ve}^+)$ and $(g_{ve}^-,g_{ve}^-)$ have two different geometric interpretations as non-degenerate geometries, which are related by a parity transformation. Especially, $(g_{ve}^+,g_{ve}^+)$ and $(g_{ve}^-,g_{ve}^-)$ imply different $\sgn V_4$ for the 4-simplex oriented volume.

For an internal face $f$, the loop holonomy $G_f(v)$ around it is given by Eq.(\ref{eq:LHsf}). The self-dual $G_f^+(v)$ and anti-self-dual $G_f^-(v)$ parts are
  \begin{equation}
    G^{\pm}_f(v)=\exp\left[\frac{\im}{2}\ep\left(\pi\sum_{e\in f}n_e\mp\sgn V_4\Theta_f\right)\hat{X}_f^{\pm}\right]
  \end{equation}
  where we have made a subdivision of the complex $\Delta$ into sub-complexes $\Delta_1\cdots\Delta_n$, such that $\sgn V_4$ is a constant on each $\Delta_i$. For the Type-A configuration in the bulk, the solution of the loop holonomy would be either $G_f(v)=(G_f^+(v),G_f^+(v))$ or $G_f(v)=(G_f^-(v),G_f^-(v))$. At these solutions the action $S_f$ becomes
  \begin{equation}
    S_{f}=2\im j_f\Phi^+, \quad \mathrm{or} \quad S_{f}=2\im j_f\Phi^-
  \end{equation}
  where $\Phi^{\pm}\equiv\ep\left(\pi\sum_{e\in f}n_e-\sgn V_4\Theta_f\right)/2$
  Explicitly, they can write down the action as
  \begin{equation}
    S_f=\im j_f\ep\left(\pi\sum_{e\in f}n_e-\sgn V_4\Theta_f\right)
  \end{equation}
  Note that $\sgn V_4$ flips sign between the two solutions $(g_{ve}^+,g_{ve}^+)$ and $(g_{ve}^-,g_{ve}^-)$.

  For an boundary triangle $f$ shared by two boundary tetrahedrons $t_{e_0}, t_{e_1}$, the holonomy is defined by Eq.(\ref{eq:LHsfB}). The self-dual $G_{e_1e_0}^+$ and anti-self-dual $G_{e_1e_0}^-$ parts are
  \begin{equation}
    \begin{split}
      G_{e_1e_0}^{\pm}&=g(n_{e_1f_e})\\
      &\times\exp\left[\frac{\im}{2}\ep\left(\pi\sum_{e\in f_e}n_e-\sgn V_4\Theta_{f_e}^B\right)\sigma_z\right]g^{-1}(n_{e_0f_e})\\
    \end{split}
  \end{equation}
 For the Type-A configuration, the solution of the loop holonomy would be either $G_{e_1e_0}=(G_{e_1e_0}^+,G_{e_1e_0}^+)$ or $G_{e_1e_0}=(G_{e_1e_0}^-,G_{e_1e_0}^-)$. In this solution the action $S_{f_e}$ becomes
  \begin{equation}
    S_{f_e}=\im j_{f_e}\ep\left(\pi\sum_{e\in f_e}n_e-\sgn V_4\Theta_{f_e}^B\right)
  \end{equation}
  Adding the asymptotic actions of internal and boundary faces together, we can get the total action
  \begin{equation}\label{eq:actionTA}
    \begin{split}
      S_f(g_{l_1l_2},n_e,\ep)&|_{\mathrm{Type-A}}=\sum_{f_i}S_{f_i}+\sum_{f_e}S_{f_e}\\
      &=-\im\ep\sgn V_4\bigg(\sum_{f_i} j_{f_i} \Theta_{f_i}+\sum_{f_e} j_{f_e} \Theta_{f_e}^B\bigg)\\
      &\quad +\im \ep\pi \bigg(\sum_{e}n_e\sum_{f\in e}j_{f}\bigg)
    \end{split}
  \end{equation}
  The action $S_f(g_{l_1l_2},n_e,\ep)|_{\mathrm{Type-A}}$ is almost the same as the non-degenerate one $S_f(g_{l_1l_2},n_e,\ep)|_{\mathrm{Non-deg}}$ in Eq.(\ref{eq:actionNondeg}). The only difference is that in the Type-A case, the action is $\gamma$ independent.

  As a result, for any Regge-like configurations $j_f$ and a Regge boundary data $n_{ef_e}$, we can have a Type-A asymptotics by summing over all Type-A degenerate critical configurations $x_c$
  \begin{equation}\label{eq:ATA}
    \begin{split}
      Z_{j_{f}}&(\Delta)|_{\mathrm{Type-A}}\sim\sum_{x_c} C(x_c)|_{\mathrm{Type-A}}\left[1+\mathcal{O}\left(\frac{1}{\lambda}\right)\right]\\
      &\times  \exp\lambda\sum_{\Delta_i}\bigg[-\im\ep\sgn V_4\bigg(\sum_{f_i} j_{f_i} \Theta_{f_i}+\sum_{f_e} j_{f_e} \Theta_{f_e}^B\bigg)\\
      &\quad +\im \ep\pi \bigg(\sum_{e}n_e\sum_{f\in e}j_{f}\bigg)\bigg].
    \end{split}
  \end{equation}

  \subsection{Type-B asymptotics}
A Type-B degenerate configuration $\{j_f,g_{ve},n_{ef}\}$ gives a so called vector geometry on the complex $\Delta$. The vector geometry is determined by the discrete geometric variables $V_f(v)$ and $V_f(e)$ which are 3-vectors. They are interpreted as the normal to the triangle $f$. Given a Type-B degenerate configurations $\{j_f,n_{ef},g_{ve}\}$, we can reconstruct them by using semi-geometrical variables $X_{ef}\equiv X_{ef}^{\pm}=\gamma j_f n_{ef}$
  \begin{eqnarray}
    V_f(e)&\equiv& 2X_{ef}\\
    V_f(v)&\equiv& h_{ve}\rhd 2X_{ef}
  \end{eqnarray}
  The same as the discussion in the non-degenerate case, because of the parallel transportation of the vector $n_{ef}$, the loop holonomy of an internal face $f_i$ and the holonomy of a boundary face $f_e$ can be written in the following way respectively
  \begin{eqnarray}
    G_{f_i}(e)&=&\exp(\im\phi_{f_i}n_{ef_{i}}\cdot\sigma)\\
    G_{f_e}(e_1e_0)&=&g(n_{e_1f})\exp(\im\phi_{f_i}\sigma_z)g^{-1}(n_{e_0f})
  \end{eqnarray}
  Thus the action becomes
  \begin{equation}
    S_{f}=2\im j_f\phi_f
  \end{equation}

For a given vector geometry variables $V_f(e)$ and $V_f(e)$, we can uniquely determined the solutions of $h_{ve}$ as $\exp(i\Phi_{ve}J)\in$SO(3). However in spin foam model what we are using is the spinor representation of SU(2) group. Because SU(2) is the double cover of SO(3), $\forall h_{ve}\in$SO(3), there are two SU(2) elements $h_{ve}^1$ and $h_{ve}^2$ with $h_{ve}^1=-h_{ve}^2$ corresponding to the same vector geometry $V_f(e)$ and $V_f(v)$. Thus $\phi_f$ is given by
  \begin{eqnarray}
      \phi_{f_i}&=\frac{1}{2}\Phi_{f_i}+\pi\sum_{e\in f_i}n_e\\
      \phi_{f_e}&=\frac{1}{2}\Phi_{f_e}+\pi\sum_{e\in f_e}n_e
  \end{eqnarray}
  where $\Phi$ is an SO(3) angle determined by the vector geometry only and $n_e=0,1$ correspond to solutions $h_{ve}^1$ and $h_{ve}^2$ respectively. 

  Then inserting the angles $\phi_{f_i}$ and $\phi_{f_e}$ back to the action $S_f$, we obtain
  \begin{equation}
      S_f|_{\mathrm{Type-B}}
      =\im\sum_{f_i} j_{f_i} \Phi_{f_i}+\im\sum_{f_e} j_{f_e} \Phi_{f_e}
   +\im 2\pi \bigg(\sum_{e}n_e\sum_{f\in e}j_{f}\bigg)
  \end{equation}
$\sum_{f\in e}j_{f}$ is an integer. So $2\sum_{f\in e}j_{f}$ is an even number. Thus when we exponentiate $S_f|_{\mathrm{Type-B}}$ to get the amplitude, the phase factor $\exp\im 2\pi \bigg(\sum_{e}n_e\sum_{f\in e}j_{f}\bigg)=1$. Thus $\exp S_f|_{\mathrm{Type-B}}$ is a function of vector geometry only. We can give a Type-B asymptotics by summing over all Type-B degenerate configurations $x_c$
  \begin{equation}\label{eq:ATB}
  \begin{split}
    Z_{j_{f}}(\Delta)|_{\mathrm{Type-B}}&\sim \sum_{x_c} C(x_c)|_{\mathrm{Type-B}}\left[1+\mathcal{O}\left(\frac{1}{\lambda}\right)\right]\\
      &\quad \times  \exp\lambda\bigg[\im\sum_{f_i} j_{f_i} \Phi_{f_i}+\im\sum_{f_e} j_{f_e} \Phi_{f_e}\bigg].
  \end{split}
  \end{equation}
  Note that if we make a suitable gauge fixing for the boundary data, we can always set $\Phi_{f_e}=0$ see e.g. \cite{Barrett:2009gg}.

  \section{General Critical Configurations and Asymptotics}\label{sec:AsymptoRegge}
\begin{figure}[htbp!]
    \centering
    \includegraphics[width=0.3\textwidth]{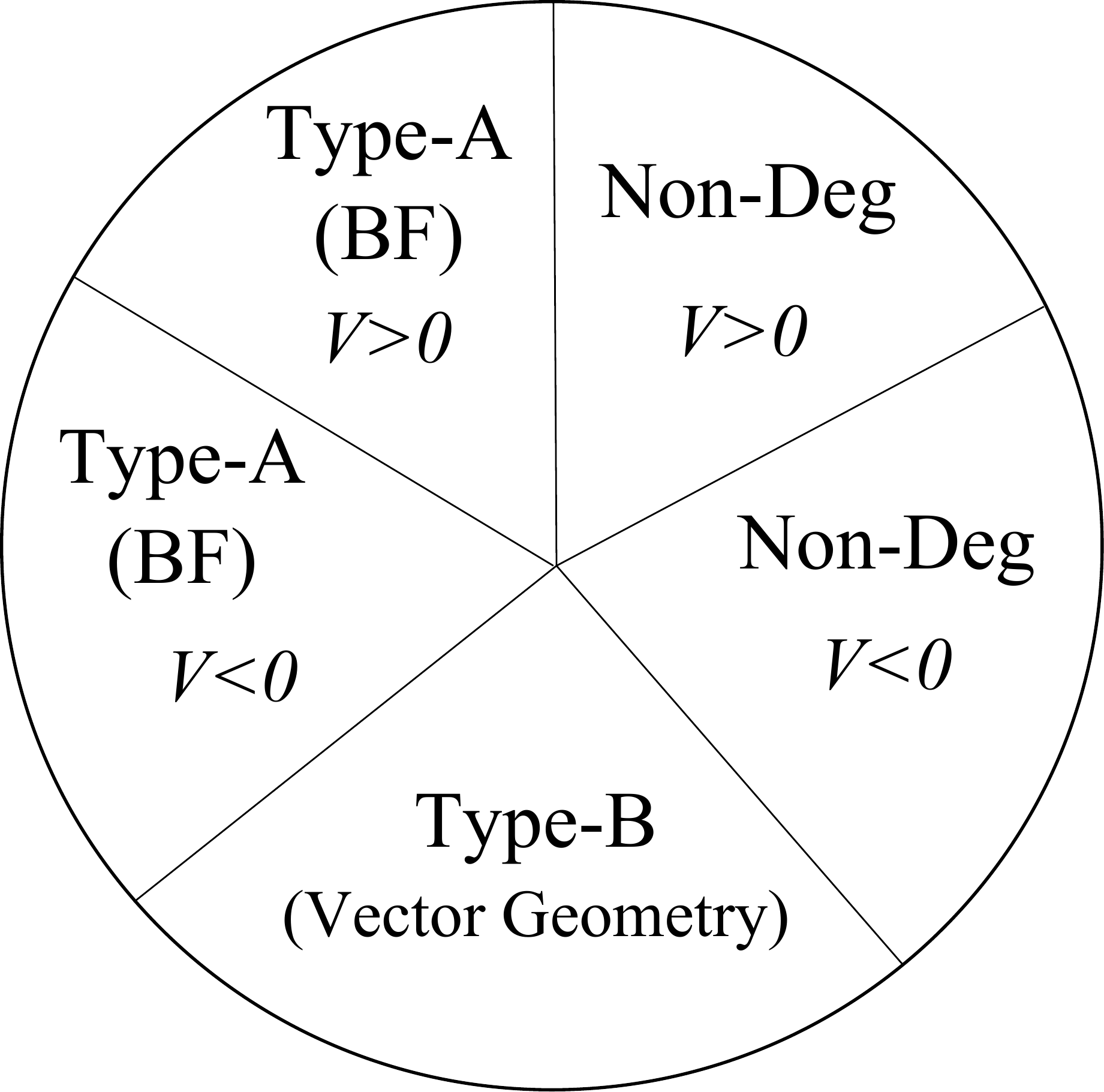}
    \caption{For a certain stationary configurations $\{j_f, g_{ve}, n_{ef}\}$, the complex $\Delta$ can be divided into 5 different types of regions: Non-Deg $V>0$, Non-Deg $V<0$, Type-A(BF) $V>0$, Type-A(BF) $V<0$, Type-B(Vector Geometry)}.\label{fig:solC}
  \end{figure}

  For a given critical configuration $\{j_f,g_{ve},n_{ef}\}$ in the most general circumstance, we can always divide the complex $\Delta$ into the non-degenerate region, Type-A degenerate region and Type-B degenerate region, according to the properties of critical configuration restricted in the regions. In non-degenerate region and Type-A degenerate region, we make further subdivision into the regions with $V_4>0$ or $V_4<0$. See Fig.\ref{fig:solC} for an illustration.

Therefore for a generic spin configuration $j_f$, the asymptotics of the spin foam amplitude $Z_{j_f}(\Delta)$ is given by a sum over all possible critical configurations $x_c$, which in general gives different subdivisions of $\Delta$ into the 5 types of regions

\begin{widetext}
\begin{eqnarray}
Z_{j_f}(\Delta)\sim\sum_{x_c}C(x_c)\left[1+o\left(\frac{1}{\lambda}\right)\right]A_{j_f}(\Delta_{\text{Nondeg}})A_{j_f}(\Delta_{\text{Deg-A}})A_{j_f}(\Delta_{\text{Deg-B}})\label{AAA}
\end{eqnarray}
where $x_c$ labels a general critical configuration $(j_f,g_{ve},n_{ef})$ admitted by the spin configuration $j_f$ and boundary data, The amplitudes $A_{j_f}(\Delta_{\text{Nondeg}}),A_{j_f}(\Delta_{\text{Deg-A}}),A_{j_f}(\Delta_{\text{Deg-B}})$ are given respectively by
\begin{eqnarray}
A_{j_f}(\Delta_{\text{Nondeg}})&=&\prod_{i=1}^{n(x_c)}\exp-i\lambda\left[\ep\ \mathrm{sgn}(V_4) \sum_{\text{internal}\ f} \gamma j_f\Theta_f+\ep\ \mathrm{sgn}(V_4) \sum_{\text{boundary }f} \gamma j_f\Theta^B_f+\pi \sum_{e}n_e\sum_{f\subset t_e} j_f\right]_{\Delta_{\text{Nondeg}},\Delta_i(x_c)}\nonumber\\
A_{j_f}(\Delta_{\text{Deg-A}})&=&\prod_{j=1}^{n'(x_c)}\exp-i\lambda\left[\ep\ \mathrm{sgn}(V_4) \sum_{\text{internal}\ f} j_f\Theta_f+\ep\ \mathrm{sgn}(V_4) \sum_{\text{boundary }f} j_f\Theta^B_f+\pi \sum_{e}n_e\sum_{f\subset t_e} j_f\right]_{\Delta_{\text{Deg-A}},\Delta'_j(x_c)}\nonumber\\
A_{j_f}(\Delta_{\text{Deg-B}})&=&\exp-i\lambda\left[\sum_{\text{internal}\ f} j_f\Phi_f+\sum_{\text{boundary }f} j_f\Phi^B_f\right]_{\Delta_{\text{Deg-B}}}
\end{eqnarray}
\end{widetext}

If we defined the physical area as $A_f=\gamma j_f$, then the Type-A action turns into
  \begin{equation}\label{eq:GTAnew}
    \begin{split}
      S_f|_{\mathrm{Type-A}}
      &=\frac{\im\ep\sgn V_4}{\gamma}\bigg[\sum_{f_i} A_{f_i} \Theta_{f_i}+\sum_{f_e} A_{f_e} \Theta_{f_e}\\
      &\quad \pm\pi \bigg(\sum_{e}n_e\sum_{f\in e}A_{f}\bigg)\bigg]
    \end{split}
  \end{equation}
 and the Type-B action turns into
  \begin{equation}\label{eq:GTBnew}
    S_f|_{\mathrm{Type-B}}
      =\frac{\im}{\gamma}\bigg[\sum_{f_i} A_{f_i} \Phi_{f_i}+\sum_{f_e} A_{f_e} \Phi_{f_e}\bigg]
  \end{equation}
  Here we consider the case when Barbero-Immirzi parameter $\gamma\ll 1$ mentioned in \cite{Bianchi:2009ri}\cite{Magliaro:Re2011}\cite{Rovelli:3P2011}. Then the Type-A degenerate parts Eq.(\ref{eq:GTAnew}) and Type-B degenerate parts Eq.(\ref{eq:GTBnew}) oscillate much more violently than the non-degenerate amplitude $A_{j_f}(\Delta_{\text{Nondeg}})$. When we sum over all spins $j_f$ to get the total spin foam amplitude, we expect that the non-degenerate critical configurations are dominating the large-$j$ asymptotics in the case of $\gamma\ll 1$. Our conjecture is suggested by the Riemann-Lebesgue lemma, which states that
  \begin{quote}
    For all complex $L^1$-function $f(x)$ on $\R$,
    \begin{equation}
      \int^{\infty}_{-\infty}f(x)\ex^{\im\alpha x}\dd x=0,\quad \mathrm{as}\quad \alpha\rightarrow\pm\infty.
    \end{equation}
  \end{quote}

  \section{Conclusion and discussion}
  In this work we study the large-$j$ asymptotic behavior of the Euclidean EPRL spin foam amplitude on a 4d simplicial complex with an arbitrary number of simplices. The asymptotics of the spin foam amplitude is determined by the critical configurations of the spin foam action. Here we show that, given a stationary configuration $\{j_f, g_{ve},n_{ef}\}$ in general, there exists a partition of the simplicial complex $\Delta$ into three types of regions: Non-degenerate region, Type-A(BF) region, Type-B(vector geometry) region. All of the three regions are simplicial sub-complexes with boundaries. The stationary configuration implies different types of geometries in different types of regions, i.e. (1) The critical configuration restricted in Non-degenerate region implies a non-degenerate discrete Euclidean geometry; (2) The critical configuration restricted in Type-A region is degenerate of Type-A in our definition of degeneracy, but it still implies a non-degenerate discrete Euclidean geometry; (3) The critical configuration restricted in Type-B region is degenerate, and implies a vector geometry.

  With the critical configuration $\{j_f, g_{ve},n_{ef}\}$, we can further make a subdivision of the Non-degenerate region and Type-A region into sub-complexes (with boundary) according to their Euclidean oriented 4-volume $V_4(v)$ of the 4-simplices, such that $\mathrm{sgn}(V_4(v))$ is a constant sign on each sub-complex. Then in each sub-complex the spin foam amplitude at the critical configuration gives an exponential of Regge action in Euclidean signature. However we should note that the Regge action reproduced here contains a sign factor $\mathrm{sgn}(V_4(v))$ related to the oriented 4-volume of the 4-simplexes. Therefore the Regge action reproduced here is actually a discretized Palatini action with on-shell connection.

  Finally the asymptotic formula of the spin foam amplitude is given by a sum of the amplitudes evaluated at all possible critical configurations, which are the product of the amplitudes associated to different type of geometries.

  We give the critical configurations of the spin foam amplitude and their geometrical interpretations explicitly. However we did not answer the question such as whether the non-degenerate critical configurations are dominating the large-$j$ asymptotic behavior in general or not, although we expect the non-degenerate configurations are dominating when the Barbero-Immirzi parameter $\gamma$ is small. To answer this question in general requires a detailed investigation about the rank of the Hessian matrix in general circumstances. We leave the detailed study about its rank to the future research.

  In this work we show that given a Regge-like spin configuration $j_f$ on the simplicial complex, the stationary configurations $\{j_f, g_{ve},n_{ef}\}$ are non-degenerate, and there is a unique stationary configurations $\{j_f, g_{ve},n_{ef}\}$ with the oriented 4-volume $V_4(v)>0$ (or $V_4(v)<0$) everywhere. We can regard the critical configuration $\{j_f, g_{ve},n_{ef}\}$ with $V_4(v)>0$ as a classical background geometry, and define the perturbation theory with the background field method. Thus with the background field method, the n-point functions in spin foam formulation can be investigated as a generalization of \cite{Bianchi:2009ri,Rovelli:3P2011} to the context of a simplicial manifold.

  \section{Acknowledgements}
  The authors would like to thank E.Bianchi, L.Freidel, T.Krajewski, S.Speziale, and C.Rovelli for discussions and communications. M.Z. is supported by CSC scholarship No.2010601003.

  \vfill

\bibliographystyle{apsrev4-1}
\bibliography{BiblioCarlo,BiblioZhang}

\end{document}